\documentclass{aa}

\usepackage[utf8]{inputenc}
\usepackage[T1]{fontenc}

\usepackage[varg]{txfonts}

\usepackage{hyperref}
\usepackage{graphicx}   
\usepackage{amsmath}    
\usepackage{amssymb}    
\usepackage{threeparttable}
\usepackage{booktabs}
\usepackage{verbatim}
\usepackage[all]{hypcap}

\begin{document}

\title{The case for two-dimensional galaxy-galaxy lensing}

\author{Andrej Dvornik\inst{1}
\and Sebastiaan L. Zoutendijk\inst{1}
\and Henk Hoekstra\inst{1}
\and Konrad Kuijken\inst{1} 
}

\institute{Leiden Observatory, Leiden University, P.O. Box 9513, 2300 RA Leiden, the Netherlands\\ \email{dvornik@strw.leidenuniv.nl}}

\date{Received ???; Accepted ???}


\titlerunning{Two-dimensional galaxy-galaxy lensing}
\authorrunning{A. Dvornik et al.}

\abstract{We revisit the performance and biases of the two-dimensional approach to galaxy-galaxy lensing. This method exploits the information for the actual positions and ellipticities of source galaxies, rather than using only the ensemble properties of statistically equivalent samples. We compare the performance of this method with the traditionally used one-dimensional tangential shear signal on a set of mock data that resemble the current state-of-the-art weak lensing surveys. We find that under idealised circumstances the confidence regions of joint constraints for the amplitude and scale parameters of the NFW model in the two-dimensional analysis can be more than three times tighter than the one-dimensional results. Moreover, this improvement depends on the lens number density and it is larger for higher densities. We compare the method against the results from the hydrodynamical EAGLE simulation in order to test for possible biases that might arise due to lens galaxies being missed, and find that the method is able to return unbiased estimates of halo masses when compared to the true properties of the EAGLE galaxies. Because of its advantage in high galaxy density areas, the method is especially suitable for studying the properties of satellite galaxies in clusters of galaxies.}

\maketitle

\section{Introduction}
\label{sec:intro}

One of the fundamental ingredients needed to understand galaxy formation is the relation between stellar mass and the host halo mass \citep[e.g.][]{Courteau2014}. However, inferring the total mass from a galaxy's emitted light is not feasible. We must instead rely on different probes to constrain the mass of dark matter haloes around galaxies one wants to study. A powerful mechanism that can be used for this is gravitational lensing,  when matter inhomogeneities deflect light rays from distant objects along their path.  As a consequence the images of distant objects (sources) appear to be tangentially distorted around foreground galaxies (lenses). The strength of the distortion is proportional to the amount of mass associated with the lenses (and consequently the dark matter haloes) and it is stronger in the proximity of the centre \citep[for a thorough review, see][]{Bartelmann1999}. 

Weak gravitational lensing induces a coherent tangential distortion. Under the assumption that galaxies are randomly oriented, the lensing signal can be inferred by simply averaging the ellipticities of the source galaxies. The typical change in ellipticity due to gravitational lensing is much smaller than the intrinsic ellipticity of the source, even in the case of clusters of galaxies. The weak gravitational lensing signal from a single galaxy halo is therefore too weak to be detected, and we must rely on a statistical approach in which   the contributions from different lens galaxies are stacked, selected by similar observational properties (e.g. stellar masses, luminosities, size). The usual method used to analyse weak lensing data is to average the tangential component of the distortion in radial bins. As the signal from a single lens is purely tangential, this is a succinct way of showing the information contained in the distortions induced by one lens; there is no information lost in azimuthally averaging a radially symmetric signal and therefore the  mass distribution of the lens can be perfectly determined from this radial profile. Average halo properties, such as halo masses, are then inferred from the resulting high signal-to-noise ratio measurements. This technique is commonly referred to as galaxy-galaxy lensing, and it is used as a method to measure statistical properties of dark matter haloes around galaxies \citep[e.g.][]{Leauthaud2011, VanUitert2011, Velander2014, Cacciato2013, Viola2015}. The stacking  mentioned here is not required per se, but it provides a convenient and unbiased data compression method that also allows for separate study of central and satellite galaxies. However, it does typically result in a loss of information about the halo properties. 

For the lenses that do not exist in isolation the signal is not purely tangential. In this case the distortions around a lens are the sum of the tangential patterns of all the neighbouring lenses. An azimuthal average of these distortions will discard the azimuthal information that is present in this case. This non-optimal use of information will result in a less precise mass estimation than would be possible with a two-dimensional method. 

Here we revisit a different method of analysing galaxy-galaxy lensing data,  first proposed by \citet{Schneider1997}, and we make a case for why it should be considered again:  it uses the unique signatures of overlapping regions of lenses to constrain the halo properties more precisely. Two-dimensional galaxy-galaxy lensing tries to fit a two-dimensional shear field directly to the galaxy ellipticity measurements. It was initially named `maximum-likelihood galaxy-galaxy lensing' after the fitting method it was first studied with \citep[e.g.][]{Schneider1997, Hudson1998, Geiger1999, Hoekstra2003, Hoekstra2004a, Han2014}. Maximum-likelihood galaxy-galaxy lensing is thus a misnomer and in principle one could use any form of fitting method to infer the desired parameters of the two-dimensional weak lensing maps, ideally using a fully Bayesian model \citep{Sonnenfeld2017}.

This method went out of fashion due to the unavailability of the galaxy grouping information that would accurately classify the galaxies as centrals and satellites \citep{Hoekstra2013book} as it was realised that these objects need to be modelled separately. Treating the galaxies as centrals and satellites in a statistical way when considering the stacked signal could be naturally accounted for with the halo model \citep{Seljak2000, Peacock2000, Cooray2002}, thus overcoming the observational shortcomings. In recent years the galaxy grouping information has become available due to the power of wide-field photometric surveys \citep[e.g. KiDS;][]{Kuijken2015, DeJong2015} complemented with spectroscopic group information \citep[from spectroscopic surveys like GAMA;][]{Driver2011, Robotham2011} that   treat the central and satellite galaxies deterministically \citep[e.g.][]{Sifon2015, Brouwer2016a}. One important advantage of the two-dimensional method is that it exploits all the information of the actual image configuration (the model predicts the shear for each individual galaxy image) using various parameters, including the galaxies' exact positions, ellipticities, magnitudes, luminosities, stellar masses and group membership information rather than using only the ensemble properties of statistically equivalent samples \citep{Schneider1997}. Moreover, the clustering of the lenses is naturally taken into account, although it is more difficult to account for the expected diversity in density profiles \citep{Hoekstra2013book}. 

The outline of this paper is as follows. In Sect. \ref{sec:mla} we present the maximum likelihood formalism used for galaxy-galaxy lensing, for both one-dimensional and two-dimensional methods. In Sect. \ref{sec:lens_model} we present the lens model used to construct the mock observations and investigation of EAGLE galaxies \citep{Schaye2015, Crain2015, McAlpine2015}. The mock observations are further described in Sect. \ref{sec:proof} where we also test the two-dimensional method and examine the limitations in the case of masked data. In Sect. \ref{sec:eagle} we examine the EAGLE simulation \citep{Schaye2015} using the two-dimensional galaxy-galaxy lensing methodology. We conclude and discuss in Sect. \ref{sec:conclusions}. Throughout the paper we use the following cosmological parameters   in the calculation of the distances and other relevant properties \citep[][as used in the EAGLE simulation]{PlanckCollaboration2014}: $\Omega_{\text{m}} = 0.307$, $\Omega_{\Lambda} = 0.693$, $\sigma_{8} = 0.8288$, $n_{\text{s}} = 0.9611$, $\Omega_{\text{b}} = 0.04825$, and $h = 0.6777$. All the measurements presented in the paper are in comoving units.

\section{2D galaxy-galaxy lensing formalism}
\label{sec:mla}

The likelihood of a model with a set of parameters $\theta$ given data $\mathbf{d}$ is parametrised in the  form
\begin{equation}
\label{eq:likely}
\mathcal{L}(\mathbf{\theta}\, \vert\, \mathbf{d}) =  \frac{1}{\sqrt{\left(2 \pi \right)^{n} \vert \mathbf{C} \vert}} \exp \left[-\frac{1}{2} \left(\mathbf{m} (\mathbf{\theta}) - \mathbf{d} \right)^{T}{\mathbf{C}}^{-1} \left(\mathbf{m} (\mathbf{\theta}) - \mathbf{d}\right) \right],
\end{equation}
where $\mathbf{m}(\theta)$ is the value of $\mathbf{d}$ predicted by the model with parameters $\theta$. We assume the measured data points $\mathbf{d} = [d_{i}, \dots, d_{n}]$ are drawn from a normal distribution with a mean equal to the true values of the data. The likelihood function accounts for correlated data points through the covariance matrix $\mathbf{C}$. The covariance matrix $\mathbf{C}$ consists of two parts, the first  arising from the shape noise and the second  from the presence of cosmic structure between the observer and the source \citep{Hoekstra2003a}:
\begin{equation}
\mathbf{C} = \mathbf{C}^{\mathrm{shape}} + \mathbf{C}^{\mathrm{LSS}} \,.
\end{equation} 
Using the Equation \ref{eq:likely}, the parameters of the best-fitting model can be determined with
\begin{equation}
\tilde{\mathbf{\theta}}(\mathbf{d}) \equiv \underset{\mathbf{\theta}}{\mathrm{argmax}}\, \mathcal{L}(\mathbf{\theta}\, \vert\, \mathbf{d}) = \underset{\mathbf{\theta}}{\mathrm{argmin}} \chi^{2} (\mathbf{\theta}\, \vert\, \mathbf{d}).
\end{equation}
For convenience we define
\begin{equation}
\chi^{2}_{\mathrm{min}} (\mathbf{d}) \equiv \chi^{2} (\,\tilde{\mathbf{\theta}}\,(\mathbf{d})\, \vert\, \mathbf{d})
\end{equation}
as the value of the chi-square statistic for the best-fitting  model, which is also the minimal value of the chi-square statistic.

When  fitting one-dimensional tangential shear profiles stacked over a sample of lenses, the likelihood function can be written as
\begin{align}\label{eq:tangential}
\mathcal{L}(M_{\mathrm{h}}, M_{\star}, c  \, \vert \, \gamma_{\mathrm{t}}^{\mathrm{obs}}) =& \\ \nonumber
= \prod_{i} \frac{1}{\sigma_{\gamma_{\mathrm{t}}, i} \sqrt{2 \pi}} &\exp \left[-\frac{1}{2} \left(\frac{\gamma_{\mathrm{t}, i} (M_{\mathrm{h}}, R, z)  - \gamma_{\mathrm{t}, i}^{\mathrm{obs}}}{\sigma_{\gamma_{\mathrm{t}}, i}}\right)^{2} \right],
\end{align}
where we  use $m_{i} = \gamma_{\mathrm{t}, i} (M_{\mathrm{h}}, R, z)$ as the model prediction given halo mass $M_{\mathrm{h}}$, radial bin $R$, and redshift of the lens $z$, and the $d_{i} = \gamma_{\mathrm{t}, i}^{\mathrm{obs}}$ as the tangentially averaged shear of a sample of lenses measured from observations. Here we also use the (statistical) uncertainty on our measurement given by the $\sigma_{\gamma_{\mathrm{t}}, i}$ calculated from the intrinsic shape noise of sources in each radial bin. Moreover, we assume that the variance $\sigma^2$ is the diagonal of the full covariance matrix
\begin{equation}
\sigma= \sqrt{\vert \mathbf{C} \vert}\,;
\end{equation}
i.e., we only account for the error due to the shape noise. Similarly, the likelihood function can be defined for the case when  fitting the two-dimensional shear field
\begin{align}\label{eq:max_like}
\mathcal{L}(M_{\mathrm{h}}, M_{\star}, c \, \vert \, \epsilon^{\mathrm{obs}}) =& \\ \nonumber
= \prod_{i} \frac{1}{\sigma_{\epsilon, i} \sqrt{2 \pi}} &\exp \left[-\frac{1}{2} \left(\frac{g_{i} (M_{\mathrm{h}}, \theta, z) - \epsilon_{i}^{\mathrm{obs}}}{\sigma_{\epsilon, i}}\right)^{2} \right],
\end{align}
where $g_{i} (M_{\mathrm{h}}, \theta, z)$ are the reduced shears evaluated at each source position $\theta$, $\epsilon_{i}^{\mathrm{obs}}$ are the observed elipticities of real galaxies, and $\sigma_{\epsilon, i}$ is the intrinsic shape noise of our galaxy sample per component, and  is the same as the $\sigma_{\gamma_{\mathrm{t}}, i}$. In practice, the two-dimensional fit to the ellipticities is carried out for each cartesian component of ellipticity $\epsilon_{1}$ and $\epsilon_{2}$ with respect to the equatorial coordinate system of the real data or mock catalogues  used in our validation study.

\subsection{Lens model}
\label{sec:lens_model}

The most widely assumed density profile for dark matter haloes is the Navarro–Frenk–White (NFW) profile \citep{Navarro1995}.
Using simple scaling relations this profile can be matched to simulated dark matter haloes over a wide range of masses and was found to be consistent with observations \citep{Navarro1995}. It is defined as
\begin{equation}
\rho_\mathrm{NFW}(r) = \frac{\delta_\mathrm{c}\, \overline{\rho}_{\mathrm{m}}}{(r/r_\mathrm{s})\, (1+r/r_\mathrm{s})^2},
\end{equation}
where the free parameters $\delta_\mathrm{c}$ and $r_\mathrm{s}$ are called the overdensity and the scale radius, respectively, and $\overline{\rho}_{\mathrm{m}}$ is the mean density of the universe, where $ \overline{\rho}_{\mathrm{m}} = \Omega_{\mathrm{m}} \rho_\mathrm{c}$ and $\rho_\mathrm{c}$ is the critical density of the universe, defined by
\begin{equation}
\rho_\mathrm{c} \equiv \frac{3H^{2}_{0}}{8 \pi G},
\end{equation}
where $H_{0}$ is the present day Hubble parameter.

The NFW profile in its usual parametrisation has two free parameters for each halo, halo mass $M_{\mathrm{h}}$, and concentration $c$, and using these parameters  is the conventional way of modelling halo profiles. However, having two free parameters for each halo is computationally very expensive. Instead,  we would like to describe these parameters through relations that depend on halo properties, and then fit to a few free parameters in these global relations instead of hundreds or thousands of free, halo-specific parameters. 

To this end, we adopt the halo mass--concentration relation of \citet{Duffy2011}, which is also an adequate description of the measured halo mass--concentration relation of central and satellite galaxies in the EAGLE simulation \citep{Schaye2015, Schaller2015} 
\begin{equation}
\label{eq:con_duffy}
c(M_{\mathrm{h}}, z) = 10.14\;  \left[\frac{M_{\mathrm{h}}}{(2\times 10^{12} M_{\odot}/h)}\right]^{- 0.081}\ (1+z)^{-1.01} \,.
\end{equation}
We also adopt the stellar mass-to-halo mass relation, as measured in the EAGLE simulation, using the functional form presented in \citet{Matthee2016},
\begin{equation}
\label{eq:smhm}
\log(M_{\star} / M_{\odot}) = \alpha - e^{\gamma} \left(M_{\mathrm{h}} / M_{\odot}\right)\,^{\beta \log(e)}\,, 
\end{equation}
where $\alpha = 11.50$, $\beta = -0.86$, and $\gamma = 10.58$.

After removing all halo-specific degrees of freedom, we introduce two new, global degrees of freedom in order to avoid recalculating the shape of the profile in every single model evaluation. They are introduced in the form of the factors $f$ and $g$, which scale the values of the scale radius $r_\mathrm{s}$ and product $\delta_\mathrm{c}r_\mathrm{s}$ relative to the values $\tilde{r}_\mathrm{s}(M_{\star})$ and $\tilde{\delta}_\mathrm{c}(M_{\star})\,\tilde{r}_\mathrm{s}(M_{\star})$ expected from a lens with a stellar mass $M_{\star}$ through the two scaling relations for $c$ and $M_{\mathrm{h}}$:
\begin{align}
r_\mathrm{s} &= f\, \tilde{r}_\mathrm{s}(M_{\star}), \nonumber \\
\delta_\mathrm{c}r_\mathrm{s} &= g\, \tilde{\delta}_\mathrm{c}(M_{\star})\, \tilde{r}_\mathrm{s}(M_{\star}).
\end{align}
Our two parameters thus correspond to a scaling of the amplitude and scale of the NFW profile. This makes the interpretation of results straightforward and is the most general parametrisation of the NFW profile. These parameters are expected to be of order unity. While the scaling relations were  measured on the EAGLE simulation, which we use to validate the method, the slight differences on exact definitions of quantities as measured on the simulations and what weak gravitational lensing infers (and scatter around the mean of those distributions) might cause slight changes in the value of the fiducial parameters. We do not expect to see any in the case of simulated, toy model observations. These lens models can be generalised to account for scatter (e.g. in the stellar mass-to-halo mass relation  or  in the concentration--mass relation)  by making it fully Bayesian, similar to the model presented in \citet{Sonnenfeld2017}.

The gravitational shear and convergence profiles are then calculated using the equations presented by \citet{Wright1999}, from which the predicted ellipticities for all the lenses are calculated according to the weak lensing relations presented in \citet{Schneider2003}. We first calculate the reduced shear for our NFW profiles,
\begin{equation}
\label{eq:red_shear}
g(\theta, z_{\mathrm{s}}) = \frac{\gamma(\theta, z_{\mathrm{s}})}{1-\kappa(\theta, z_{\mathrm{s}})}\,, 
\end{equation}
from which the ellipticities are calculated according to 
\begin{equation}
\epsilon =
        \begin{cases}
            g & \vert g \vert \leq 1 \\
            1/g^{*} & \vert g \vert > 1
        \end{cases}\,, 
\end{equation}
where we  assume that the intrinsic ellipticities of the sources average to 0, due to their random nature. In practice, we avoid the strong lensing regime by removing these sources from our catalogue.

\section{Proof of concept}
\label{sec:proof}

We created the mock catalogues in a semi-empirical manner. In order to test the method on a realistic dataset, the mock catalogues were made to  closely resemble the Kilo Degree Survey (KiDS) properties \citep{DeJong2015, Kuijken2015}. We randomly placed 30\,700 sources at a redshift of 0.7 in a 1 deg$^{2}$ field. This corresponds to the size of one KiDS tile with the number of sources reflecting the observed number density \citep{Hildebrandt2016} at the median redshift for the whole survey. 

We did not assign any intrinsic orientation or ellipticity to our sources; this uncertainty can be accommodated for directly in our maximum likelihood fits by scaling the covariance matrix (or in this case the variance used in the likelihood functions) so that the intrinsic source ellipticity uncertainty is representative of the shape noise in the KiDS survey, considering the overlap with the Galaxy and Mass Assembly survey \citep[GAMA; ][]{Liske2015}.

The generated source field was then used to calculate the weak lensing effect of the foreground lenses that we placed in the same field. We calculated the effect of each lens according to the model  presented in Sec. \ref{sec:lens_model}, using only one stellar mass for all the lens galaxies placed in the mock catalogue. We decided to assign a stellar mass of $M_{\star} = 10^{12} M_{\odot}$ and $f = g = 1$. We positioned all the lenses at the same redshift of 0.2, which is around the median redshift of the GAMA survey commonly used in KiDS galaxy-galaxy lensing studies \citep[][amongst others]{Viola2015, Sifon2015, vanUitert2016, Brouwer2016, Dvornik2018}.  The contributions from multiple lenses to the shear (and consequently ellipticity) of one source galaxy can be summed together linearly, i.e.
\begin{equation}
\label{eq:g_sum}
\gamma (\theta, z_{\mathrm{s}})_i = \sum_{j} \gamma(\theta, z_{\mathrm{s}})_{ij} \,,
\end{equation}
where the sum goes over the $j$ lenses in the catalogue, with shear evaluated at each source position $i$.\footnote{We first calculate the $\gamma (\theta, z_{\mathrm{s}})$ and the $\kappa(\theta, z_{\mathrm{s}})$, then use  Equation \ref{eq:red_shear} to calculate the reduced shear.} This means that we actually allow for contributions of neighbouring haloes, which will become evident later on in the paper. We also assume that each lens galaxy is exactly at the centre of its dark matter halo, ignoring the possibility of miscentring. When placing the lenses in our mock field, we draw their positions in the same way as for the sources, but we do not allow for exact spatial overlap of any lens. The number of lenses that we add to the KiDS-like field varies between 1 and 720 (the latter reflects the typical density of the GAMA galaxies) in order to study the performance of the method as a function of galaxy number density so we can test the effects of the neighbouring haloes.

When working with ground-based observations we would never have included strongly lensed sources in the analysis as in the majority
of   cases these sources are  also blended with the lens galaxy (the typical Einstein radius for GAMA galaxies is smaller than 5 arcsec); the simulated mock observations can have sources that are strongly lensed because we  distributed them randomly. To eliminate this problem, all the sources with $\vert g \vert > 0.3$ were removed from the catalogue in order to limit our analysis to the weak lensing regime. This threshold is quite low, but it makes sure we  always stay in the weak lensing regime that motivates the use of Equation \ref{eq:g_sum}.

Using these mock catalogues, we test our lens model and compare the results obtained using the one-dimensional stacked tangential shear method against the two-dimensional method that uses measured ellipticities directly. At the same time, this allows us to study the two methods under known conditions and makes the results easier to understand. 

The main question we want to address here is how the effective lens galaxy density influences the performance of the two-dimensional galaxy-galaxy lensing method as the unique signatures caused by the spatial lens configuration on the shear field  result in information gain for the inference of halo masses and halo concentrations.

The second question we want to address is the sensitivity of the two-dimensional method to incompleteness in the lens sample. This bias can be induced by lenses outside of the observed field (or masked from the data), and corrections to account for this effect were already studied in the past \citep{Hudson1998}. We applied a typical KiDS data mask to the generated mock catalogues and also studied the unmasked mock catalogues, but purposefully ignored a number of lenses that are present in the field.

\begin{figure*}
        \centering
        \begin{minipage}[b]{\columnwidth}
        \includegraphics[width=\columnwidth]{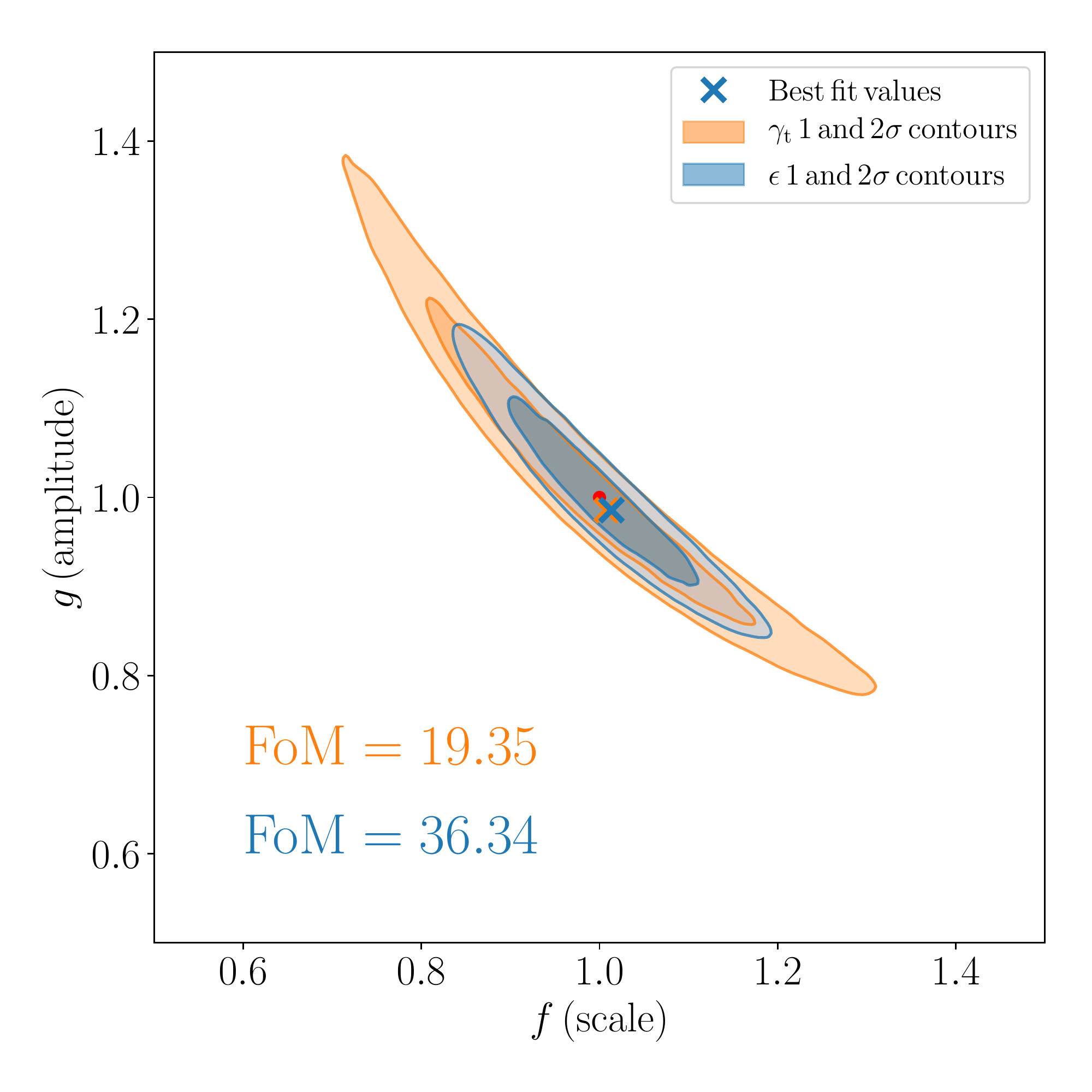}
        \end{minipage} \hfill 
        \begin{minipage}[b]{\columnwidth}
        \includegraphics[width=\columnwidth]{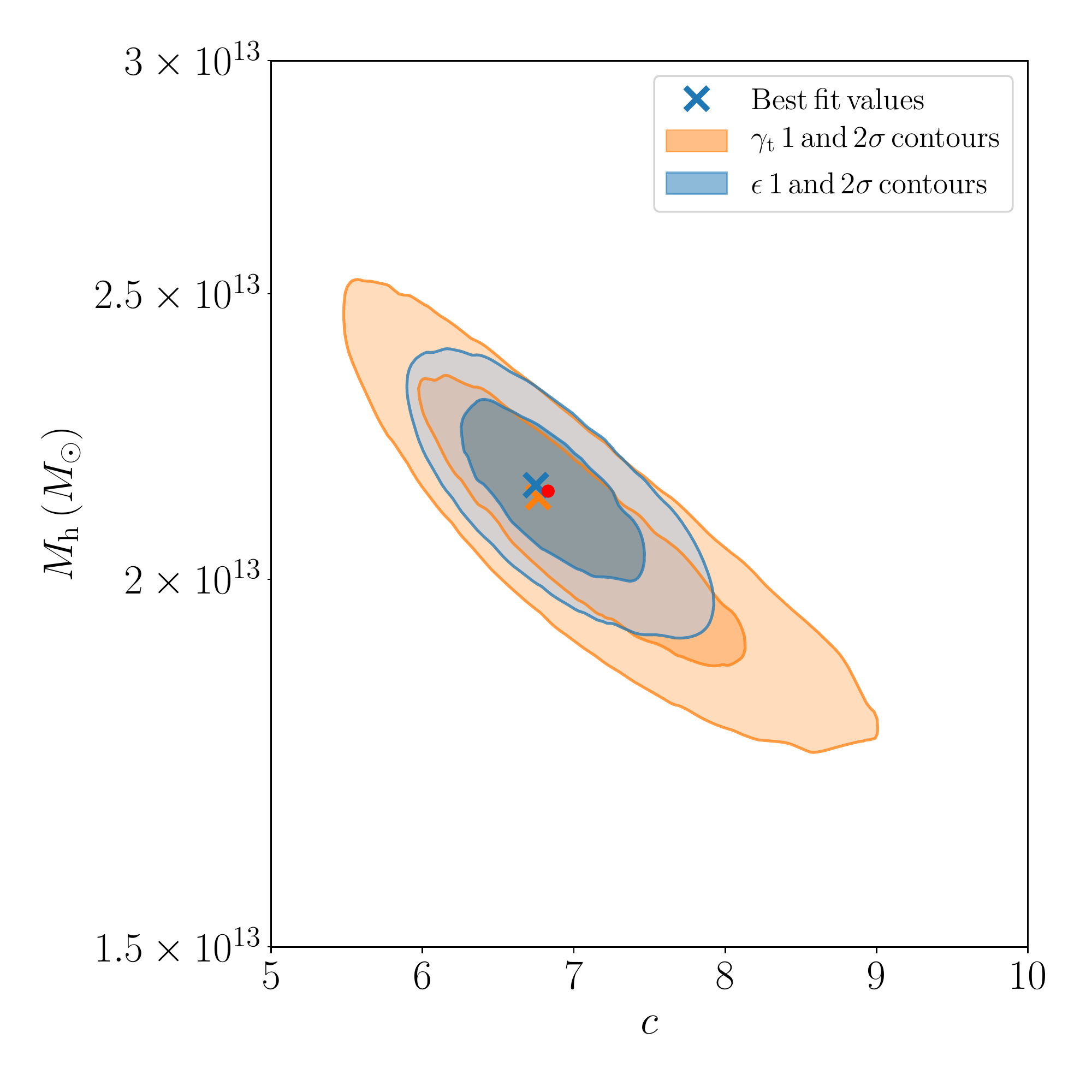}
        \end{minipage}
        \caption{Confidence areas of the scale parameter $f$ and amplitude parameter $g$ (\textit{left panel}) and the confidence areas of the halo mass $M_{\mathrm{h}}$ and halo concentration $c$ jointly derived from the constraints on the $f$ and $g$ parameters (\textit{right panel}) for an analysis of the mock KiDS+GAMA  area. Orange contours show the maximum likelihood fit on the stacked tangential shear profiles, and the blue contours the maximum likelihood fit as it was performed on the ellipticities of sources used directly, using all the galaxies in the mock field simultaneously. Shown are  the   best-fitting values for each method (orange
and blue crosses) and   the fiducial lens model (red circle). The contours correspond to the case with 50 lenses per deg$^2$ in the simulated field.}
        \label{fig:contours1}
\end{figure*}

We first applied both methods to the same source sample in which we varied the number of lenses from 1 to 720 deg$^{-2}$. We assigned the uncertainty of the measured shapes to $\sigma_{\epsilon} = 0.3$. For the one-dimensional method, each source's uncertainty was further weighted by $\sqrt{N}$, where $N$ is the number of lenses that contribute to the total shear of that source, to account for proper covariance between the sources. This is naturally captured by the two-dimensional galaxy-galaxy lensing method. Furthermore, we limited ourselves to a subset of sources that we use in both methods. The subset of sources is selected by the smallest and largest annuli ($R_{\mathrm{max}}$) in which we calculate the tangentially averaged shear profiles of our galaxies. This allows us to directly compare the methods, as for the case of one lens. Given that we use the same source galaxies, the results from the two methods should be exactly the same. At the same time, we also select the lens galaxies that are at least $R_{\mathrm{max}}$  from the field edge to minimise the effects of missing source galaxies beyond our simulated field. We fit the data using Equations \ref{eq:tangential} and \ref{eq:max_like} for the tangential shears measured on the data and the ellipticities as created in our mock catalogues, respectively. In the fit we vary the parameters $f$ and $g$, which scale the reference NFW profile for the typical scale and amplitude. We sample the values of $f$ and $g$ on a Latin hypercube grid \citep{McKay1979} using $500$ points. We compare the inferred best-fit values and the $1\sigma$ and $2\sigma$ contours obtained from a $\chi^2$ surface, which is in turn computed from the aforementioned grid using a interpolation on a finer linearly spaced grid. Using this information we calculate a figure of merit (FoM) which is defined as an inverse of the 68\% confidence level area and we study the ratio of the FoM between the one-dimensional stacked tangential shear method and the two-dimensional method.

\begin{figure*}
        \centering
        \begin{minipage}[b]{\columnwidth}
        \includegraphics[width=\columnwidth]{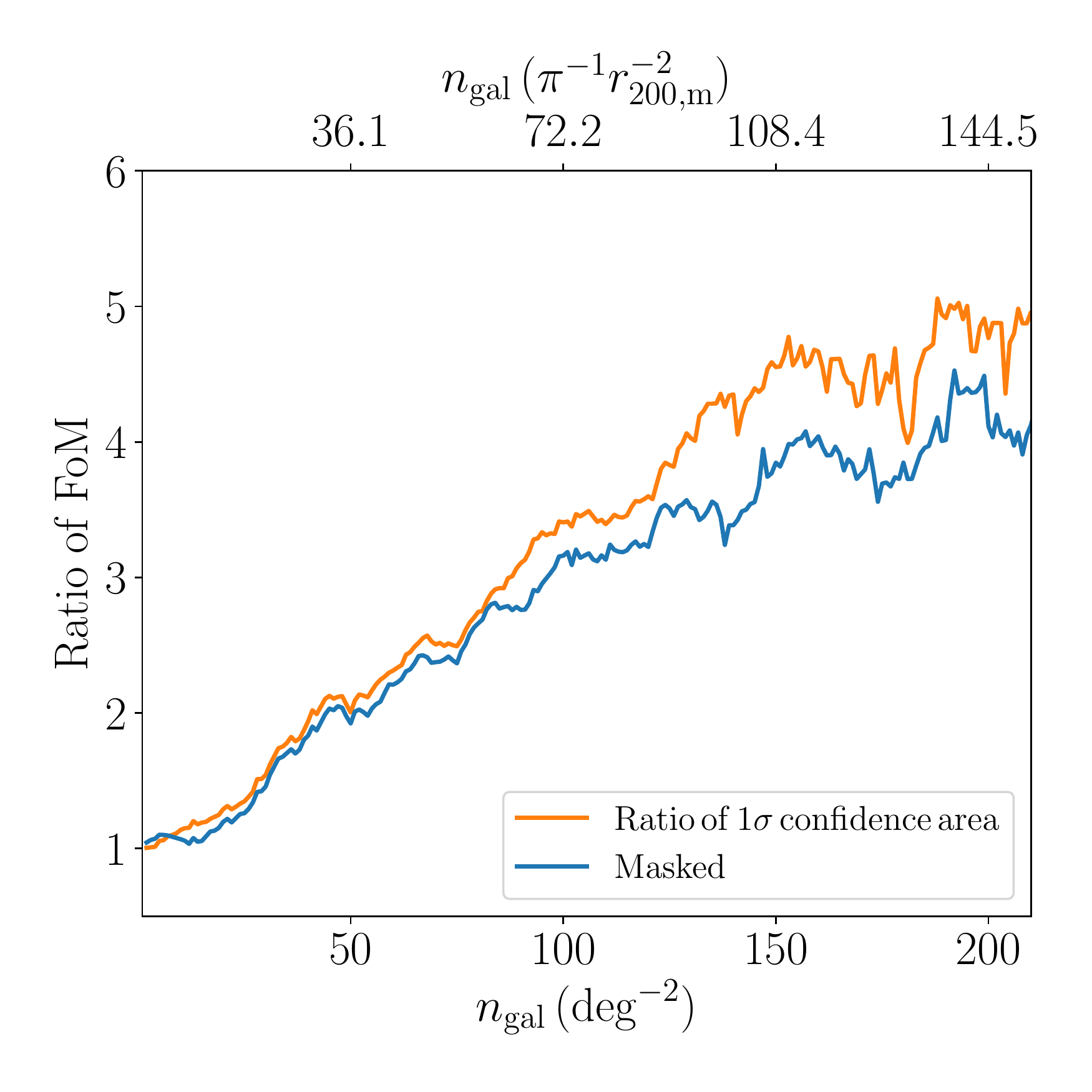}
        \end{minipage} \hfill 
        \begin{minipage}[b]{\columnwidth}
        \includegraphics[width=\columnwidth]{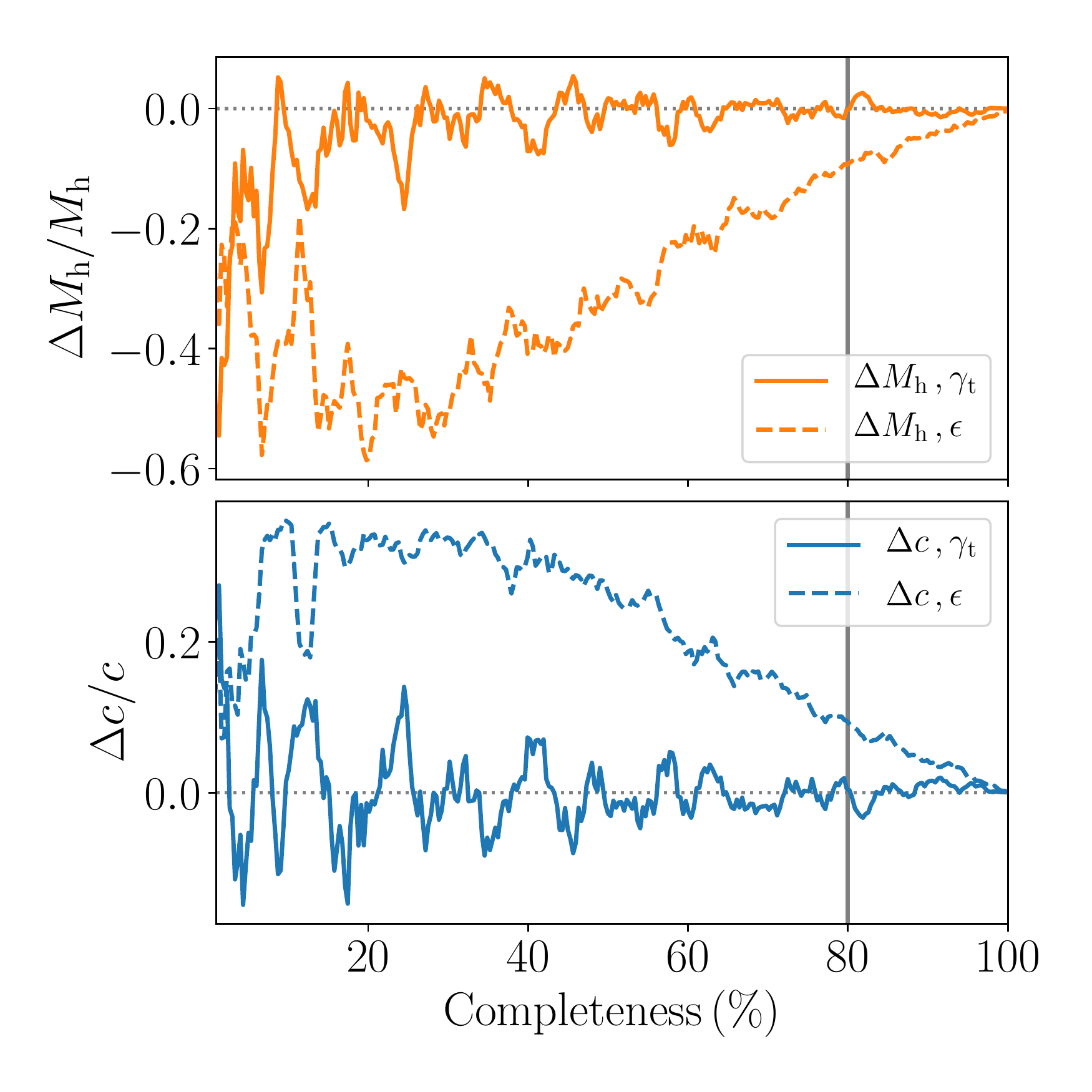}
        \end{minipage}
        \caption{\textit{Left panel:} Figure of merit  as a function of lens number density in a simulated 1 deg$^2$ field. The orange line shows the case where we consider all the galaxies in the
field, and thus gives us an estimate of improvement in precision when using a two-dimensional method. The improvement levels off at a value of around 5, which indicates that in dense galaxy fields,  the loss of signal-to-noise ratio due to the limited number of sources cannot be overcome. The blue line shows the case where we apply a typical KiDS survey mask to our mock catalogues. \textit{Right panel:} Relative shift of the halo mass $M_{\mathrm{h}}$ (\textit{top}) and halo concentration $c$ (\textit{bottom}) derived from the constraints on the $f$ and $g$ parameters from the fiducial model as a function of completeness. Shown are the shift of the recovered parameters for the one-dimensional method (solid lines) and  the shift of the recovered parameters for the two-dimensional method (dashed lines). Also shown is a typical completeness due to a mask in a KiDS like survey (vertical grey line).}
        \label{fig:ratio}
\end{figure*}

The results using 50 lenses per deg$^2$ can be seen in Fig. \ref{fig:contours1}, where we show the fiducial value of the $f$ and $g$ parameters, the best-fit values,  and the $1\sigma$ and $2\sigma$ uncertainty contours on the derived best-fit values. Similarly, in Fig. \ref{fig:contours1}, we show the constraints on the halo mass $M_{\mathrm{h}}$ and halo concentration $c$, as derived from the constraints on parameters $f$ and $g$. The best-fit values with the individual 68\% confidence intervals are listed in Table \ref{tab:fit_results}, for  the parameters $f$ and $g$, and for the halo mass $M_{\mathrm{h}}$ and halo concentration $c$. Both methods are capable of recovering the input values. What is more, the contours for the two-dimensional method are noticeably smaller. This can be   seen more clearly in Fig. \ref{fig:ratio} where the orange line shows the FoM as function of number of lens galaxies in our mock field. This figure  shows that  information is gained as the contributions of neighbouring dark matter haloes leave unique shear configuration signatures that can only be accounted for using a two-dimensional galaxy-galaxy lensing method. At low lens densities we expect the two methods to perform identically (with FoM ratio $ = 1$) as the separation of the galaxies is large enough for us to  assume that the lenses are isolated, such that $\gamma_{\mathrm{t}}$ contains all lensing information. The same effect (ratio of FoM $ = 1$) should also be observed if two lenses are exactly on the same line of sight. We note that we   consider here a noiseless mock dataset, with shape noise accounted through the covariance matrix, and this means that the signal-to-noise ratio at low densities does not influence our ability to constrain contribution of individual haloes, and consequently allows us to obtain the ideal case of FoM $ = 1$ for the case of one lens. The figure of merit stays close to 1 as long as the separations are large enough for contributions of neighbouring lenses to remain sufficiently low. With large lens galaxy number densities the lenses start overlapping,  we gain  less information, and the figure of merit starts levelling off. This is caused by  the source number density that stays the same for any number of lens galaxies we add to the field, which limits the available signal-to-noise ratio of the measured source ellipticities. 

\begin{table}
        \caption{Best-fit values.}
        \label{tab:fit_results}
        \centering
        \begin{tabular}{lcccc} % four columns, alignment for each
                \toprule
                & $f$ & $g$ & $M_{\text{h}} [10^{13} M_{\odot}]$ & $c$ \\
                \midrule
                1D&$1.0\pm^{0.14}_{0.19}$&$0.98\pm^{0.22}_{0.11}$&$2.13\pm^{0.37}_{0.43}$&$6.76\pm^{2.26}_{1.28}$ \\ \\
                2D&$1.01\pm^{0.08}_{0.11}$&$0.98\pm^{0.11}_{0.07}$&$2.15\pm^{0.24}_{0.25}$&$6.75\pm^{1.18}_{0.86}$ \\
                \bottomrule
        \end{tabular}
        \tablefoot{Best-fit values for the $f$ and $g$ parameters and for the halo mass $M_{\mathrm{h}}$ and halo concentration $c$, together with their individual 68\% confidence intervals for the model using 50 lenses per deg$^2$ in mock KiDS and GAMA like data.}
\end{table}

\begin{figure}
        \includegraphics[width=\columnwidth]{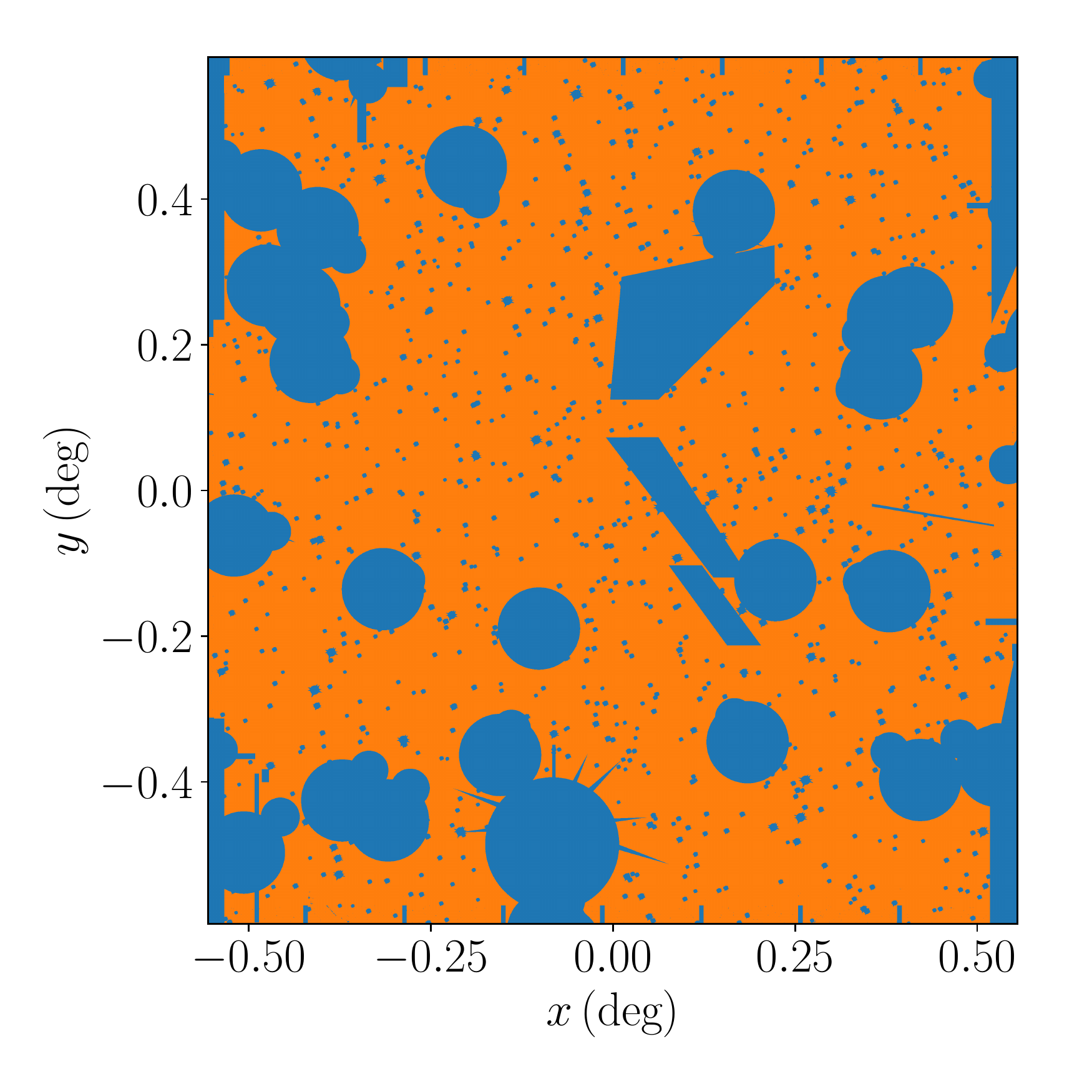}
        \caption{Typical KiDS survey \emph{r}-band mask used to evaluate the effect of masking on the inference of best-fit parameters, and the possible bias masking might introduce.}
        \label{fig:mask}
\end{figure}

We now  focus on the second question in our investigation,  whether there is any bias introduced when not all lensing galaxies are accounted for. To this end, we use the same KiDS-like mock field with $720$ lens galaxies, but now we remove one lens in each iteration, thus effectively accounting for the possible bias we might introduce in real observations by not accounting for galaxies just outside of our observed field or not accounting for lens galaxies that were masked out of the data. Figure \ref{fig:ratio} shows the shift of the best-fit parameters away from the fiducial model as a function of the field completeness, averaged over five different realisations of the lens distribution. What  is immediately clear is that the NFW fit to the one-dimensional tangential shear profiles recovers the true input parameters (as it is essentially removing any configuration information from the sample by the tangential averaging), also for the cases of low completeness. The two-dimensional method can only do this successfully at high completeness values;  any small deviation and unexpected features in the field caused by the presence of lenses  not accounted for drives the recovered values of the input parameters away from the truth as the model tries to accommodate the missing lenses.

We also study the effect of the masking introduced by a realistic KiDS survey mask, shown in Fig. \ref{fig:mask}. We apply this mask to our mock catalogues and repeat the fitting of our model to the lenses and sources that remain in the mock catalogues. We again change the number of lenses in the field and the results of this exercise can be seen in Fig. \ref{fig:ratio} (blue line). What can be observed is that the two-dimensional method, even in the case of masking, is still more precise, and that the difference in precision is a direct result of the amount of masked area. The accuracy of the method due to masking behaves in a similar way to that  shown in Fig. \ref{fig:ratio}, but the observed bias is smaller because a larger number of lenses remain in the field. A typical KiDS survey mask reduces the number of lenses by about 20\% \citep{Kuijken2015, DeJong2015, Hildebrandt2016}, which can bias the fitted parameters up to 10\%, as shown in Fig. \ref{fig:ratio} (vertical grey line). This needs to be accounted for in an application to real data.

Thus far we have ignored systematic biases in the galaxy shape measurements. They can be split into a multiplicative bias, which leads to an overall scaling of the signal, and an additive bias  that manifests as a preferred orientation of galaxies. As the former simply scales the signal, the impact on  the one-dimensional and  two-dimensional analyses is the same. The situation is different in the case of additive bias: a constant signal will simply vanish when we consider the azimuthally averaged tangential shear (in the limit of no edge effects). Even a spatially varying additive bias is expected to vanish because it typically does not align with the line connecting the lens and the source. In contrast, in the two-dimensional case we expect the $\chi^{2}$ to become poor as the systematic signal contributes to it. To examine whether this has any impact on the recovered model parameters we mimic a systematic shape measurement error by adding a constant uniform shear to our mock dataset and repeat  our analysis. We find that the overall $\chi^{2}$ surface indeed becomes offset by a constant (positive) value; however, we are nonetheless able to recover the input parameters exactly as in our fiducial case.\footnote{This can be explicitly seen by writing out the $\chi ^{2}$ with the added constant shear, $\chi ^{2} \propto (\gamma_{\text{t}} + c - m)^{2}$, where $c$ is the constant shear and $m$ the model prediction. The cross terms in the expanded form average to 0 and we gain a constant term $c^2$, which worsens the overall $\chi^{2}$, but does not inhibit the ability to minimise the $(\gamma_{\text{t}} - m)^2$ difference.}

\section{Evaluation of the two methods with the EAGLE simulation}
\label{sec:eagle}

\begin{figure}
        \includegraphics[width=\columnwidth]{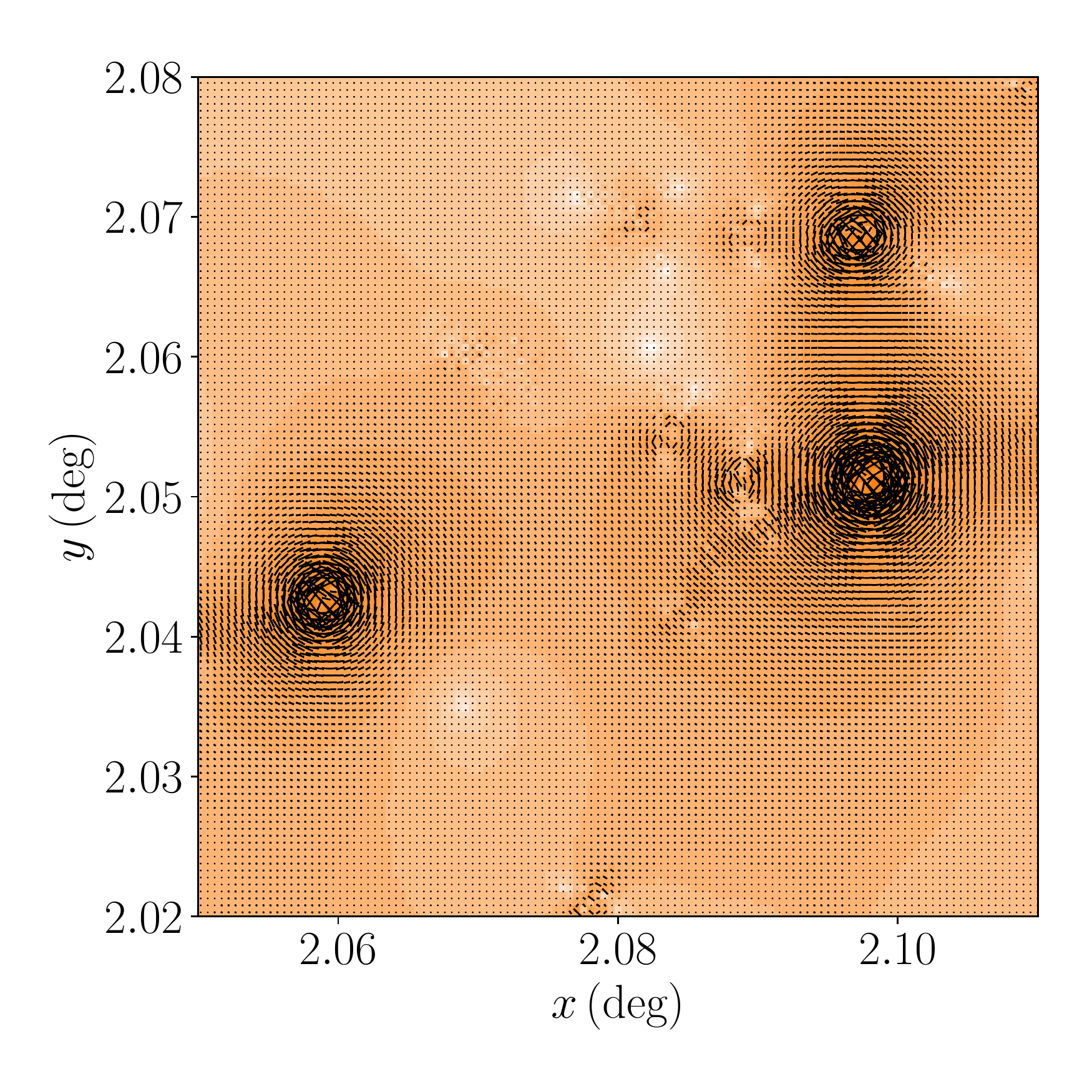}
        \caption{Segment of the shear map derived from the EAGLE particle data. A number of notable features of weak lensing are visible in this plot. The ellipticites are tangentially aligned with the lenses and the strength of gravitational lensing diminishes with distance from the lens. The lens configuration creates a unique pattern that contains information about mass distribution that is otherwise lost when tangentially averaging the observed shears.}
        \label{fig:shear_map}
\end{figure}

Motivated by the success of the two-dimensional galaxy-galaxy lensing method from the previous section, we  now focus on more realistic tests using the EAGLE hydrodynamical simulation \citep{Schaye2015, Crain2015, McAlpine2015} as our input data. Studying a simulation gives us the ability to compare our two-dimensional galaxy-galaxy lensing results against the truth,  properties as measured directly from particle properties in the simulation. We note that for the purpose of this study we do not use a lightcone generated from the EAGLE simulation. Although we include complexities of neighbouring galaxies, we do not capture projections along the line of sight or missing galaxies, for example. We use the AGN simulation \texttt{AGNdT9L0050N0752}, which has $752^3$ dark matter particles and a box size of 50 comoving Mpc \citep{Schaye2015}, and it is calibrated in such a way that it reproduces global observables of our Universe. The EAGLE simulation was also shown to correctly predict the galaxy-galaxy lensing signal when compared to the KiDS+GAMA data \citep{Velliscig2016}, for both central and satellite galaxies. We take the full particle information in a box with a comoving size of 50 Mpc which is then binned to $8195 \times 8195 \times 8195$ pixels. The box is then projected along the axes, yielding three different mass maps of the EAGLE simulation. To calculate the shear at each location, we first position the mass map at redshift of 0.2 (by scaling it comovingly), calculate the density map using the mean density of the Universe, and use the \citet{Kaiser1993} prescription to calculate the shear, rolling the edges of the map. Because we want the sources to be positioned at a redshift that resembles the typical source redshift in the KiDS survey, we calculate the convergence map using the critical surface mass density at redshift of 0.7. A small portion of an EAGLE shear map is shown in Fig. \ref{fig:shear_map}. The whole EAGLE map corresponds to a 60 deg$^2$ patch of sky.

Further information about the properties of the lens galaxies are queried from the public EAGLE database \citep{McAlpine2015}, such as the total halo mass, centre of mass, centre of potential, stellar mass, stellar mass within certain aperture, group memberships, and group properties. From the database we select galaxies with stellar masses ranging from $10^{9.6} M_{\odot}$ to $10^{11.2} M_{\odot}$, which is a range that allows us to have enough galaxies in finer stellar mass bins (which we will use for our fiducial stacked tangential shear method) and ensures that the galaxies in EAGLE are well defined in terms of simulation particle mass. From this selection of galaxies we take both the centrals and satellite galaxies. Inclusion of satellite galaxies in the study is crucial for the two-dimensional method, as otherwise  the results  can be substantially biased (up to 10\% for typical survey masks), as demonstrated in the previous section and in Fig. \ref{fig:ratio}, and at the same time it allows  their properties to be studied,  as was previously done using the one-dimensional method,  by \citet{Sifon2015}, amongst others. In total, after applying all the selection criteria, we are left with 859 galaxies (520 centrals and 339 satellites).

\begin{figure}
        \includegraphics[width=\columnwidth]{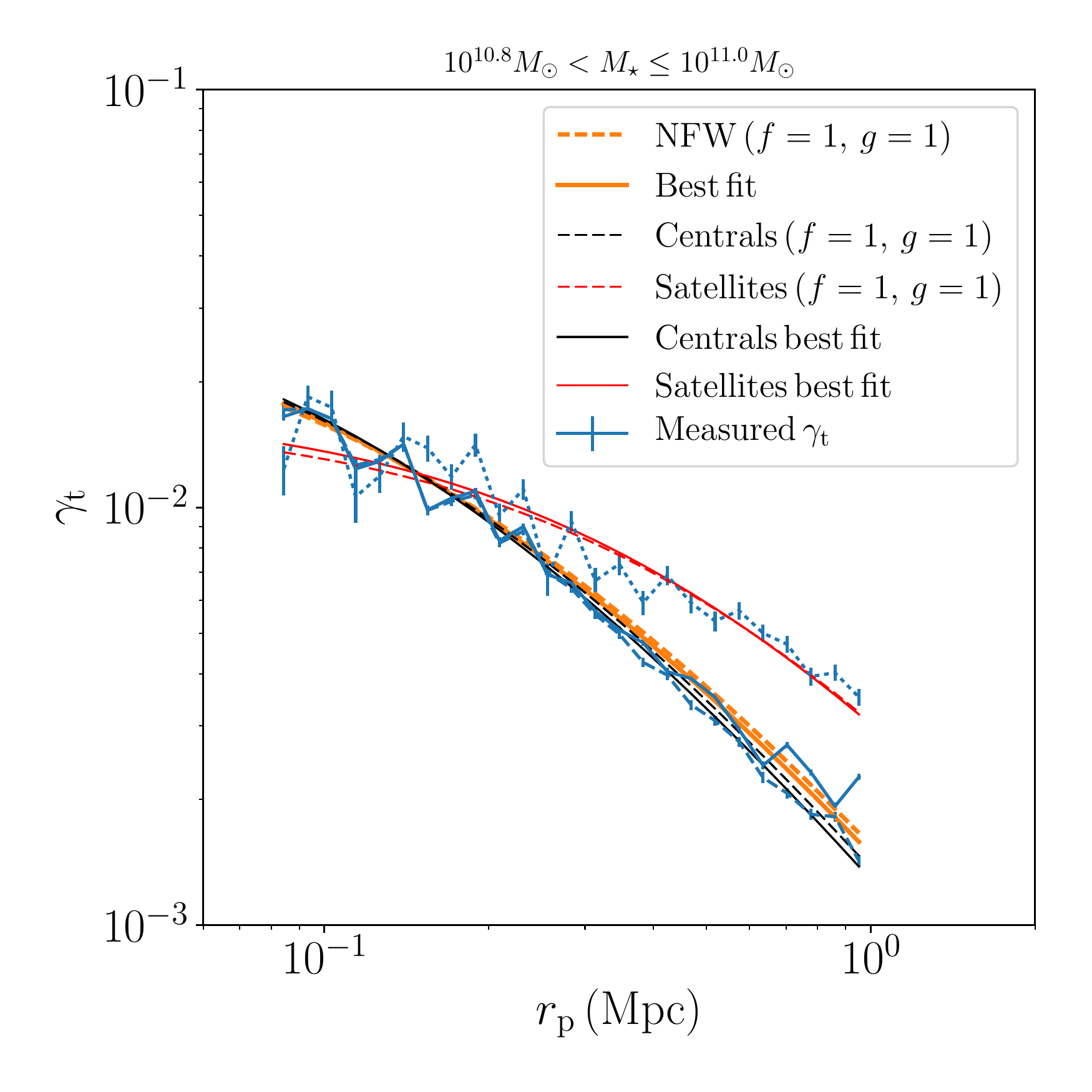}
        \caption{Stacked tangential shear profiles for the lenses selected from the EAGLE simulation (blue lines) in the $10^{10.8}$ to $10^{11.0} M_{\odot}$ stellar mass bin. The dashed orange, red, and black lines  show the signals  predicted by our fiducial lens model using $f=1$ and $g=1$ for NFW, satellites, and centrals, respectively. The total best-fitting model is shown with the solid orange line and the best-fitting model for centrals and satellites in solid black and red lines, respectively. The corresponding halo masses and concentrations of input models and the best-fit results are listed in Table \ref{tab:cm}.}
        \label{fig:gamma_eagle}
\end{figure}

We calculate the tangential shear signal for each galaxy in our sample using the tangential $\epsilon_{\rm t}$ component of the source's ellipticity around the position of the lens. The azimuthal average of the tangential ellipticity is then our unbiased estimate of the tangential shear. For the two-dimensional method, we use the $\epsilon_{1}$ and $\epsilon_{2}$ values directly. The tangential shear profiles and their averages for the $10^{10.8}$ to $10^{11.0} M_{\odot}$ stellar mass bin can be seen in Fig. \ref{fig:gamma_eagle}. The noisiness of the the individual profiles can be directly attributed to the fact that we are azimuthally averaging the data on a square grid.

\begin{figure}
        \includegraphics[width=\columnwidth]{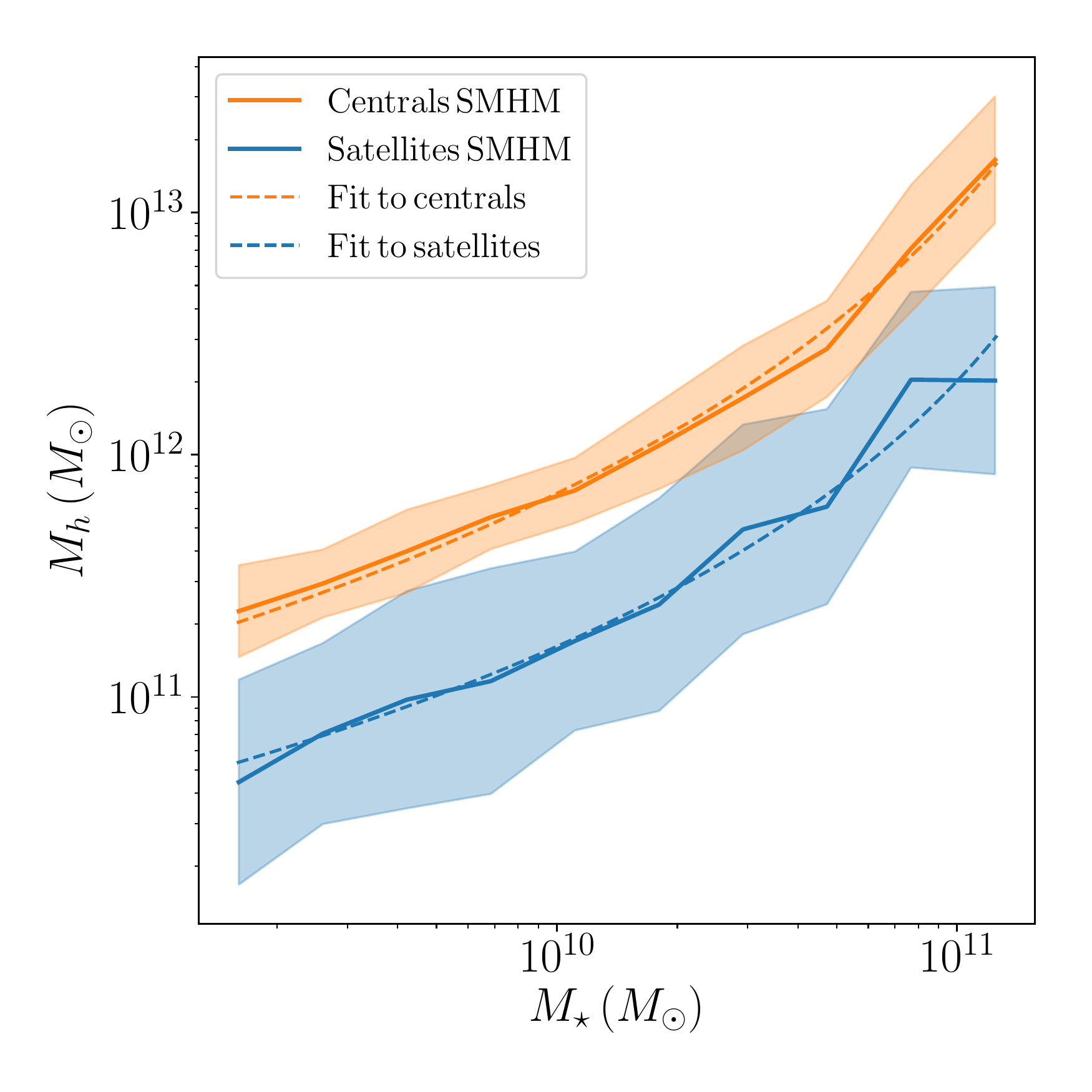}
        \caption{Stellar-to-halo mass relation for the central galaxies in the EAGLE simulation (in orange) and for the satellites (in blue). Shown are the median stellar-to-halo mass relation (solid lines) with the corresponding scatter (filled areas). The dashed  lines are the models for the stellar-to-halo mass relation   used in the analysis.}
        \label{fig:smhm}
\end{figure}

The central and satellite galaxies in the EAGLE simulation follow a different stellar-to-halo mass relation, and we also account for this in the model. For this we use the same relation  used in Sec. \ref{sec:lens_model} and we fit it to the halo masses derived from the NFW fits to the convergence field of individual galaxies, both centrals and satellites (this is done in order to use the same definition of the halo mass for both galaxy types). The two stellar-to-halo mass relations are shown in Fig. \ref{fig:smhm}, and the parameters obtained are listed in Table \ref{tab:smhm}. The different stellar-to-halo mass relations are then accounted for in the modified lens model that differentiates between the central and satellite galaxies for  the one-dimensional method and for the two-dimensional method. From the same fit we note that the concentrations of the haloes are generally lower than the prediction from \citet{Duffy2011}. While they still follow the same trend, the normalisation of the relation is lower for both centrals and satellites, with a normalisation of $0.6$ for centrals and $0.25$ for satellites, with the lower values arising because we include all the haloes,  not only the relaxed ones. This is consistent with what was found by \citet{Viola2015}. 

\begin{figure*}
        \centering
        \begin{minipage}[b]{\columnwidth}
        \includegraphics[width=\columnwidth]{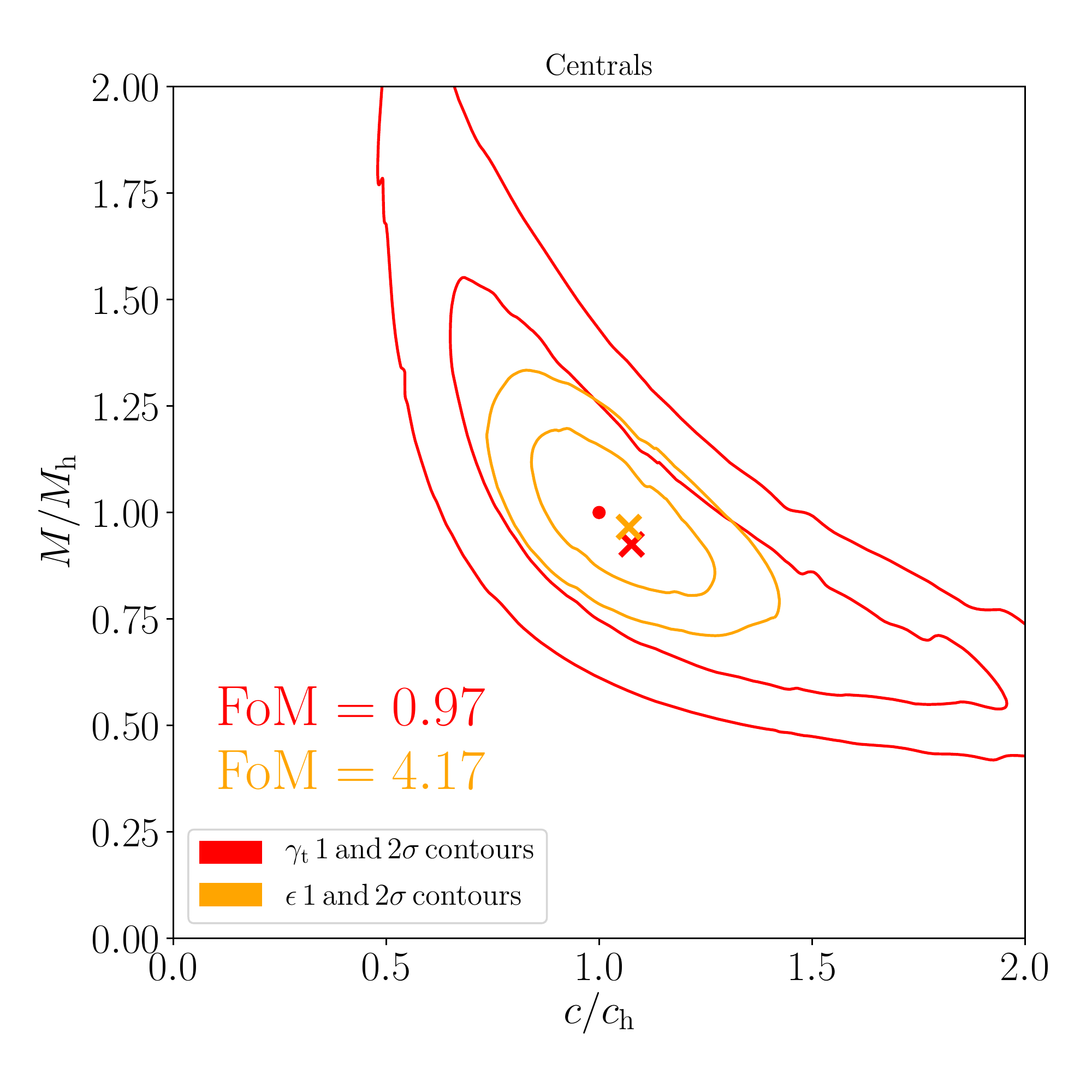}
        \end{minipage} \hfill 
        \begin{minipage}[b]{\columnwidth}
        \includegraphics[width=\columnwidth]{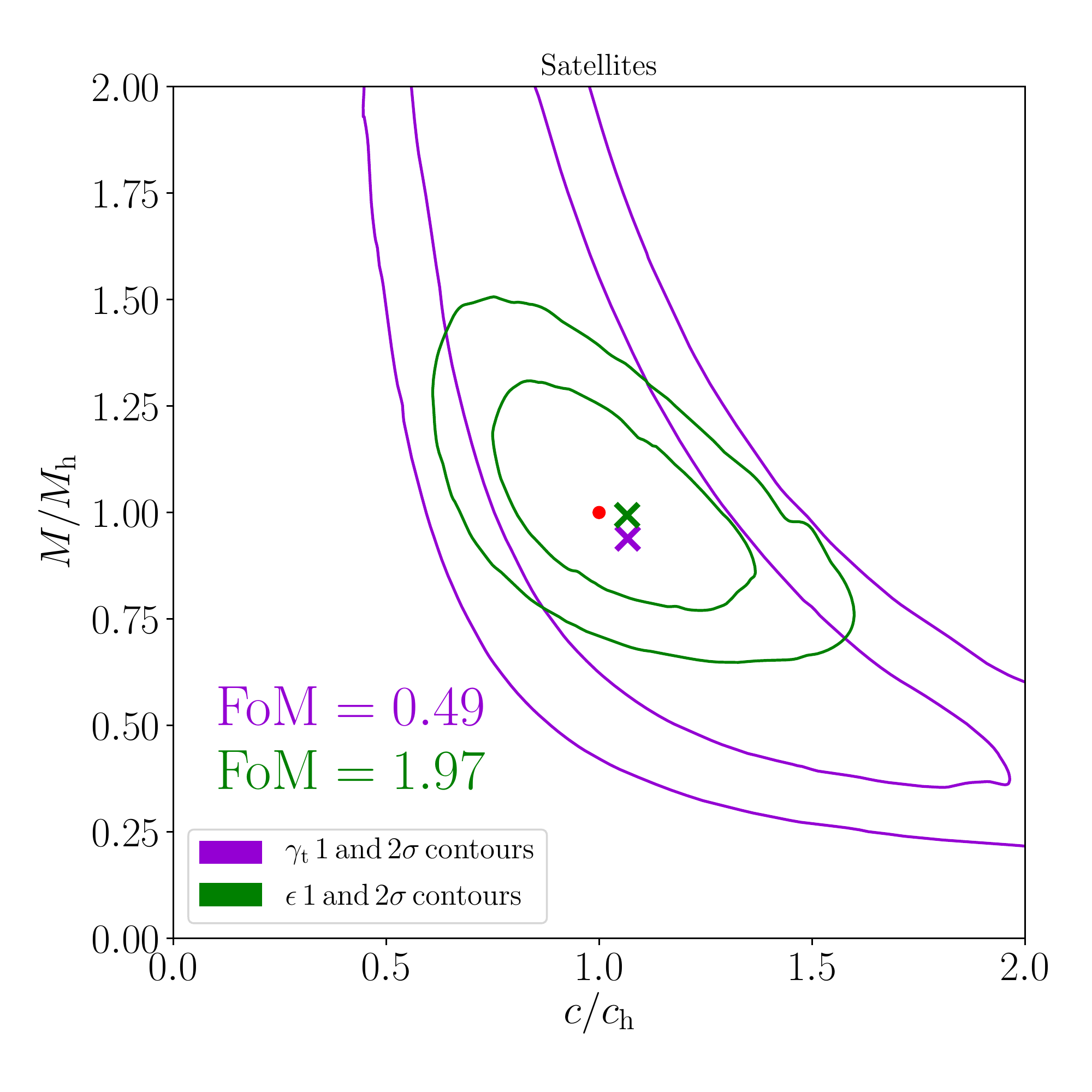}
        \end{minipage}
        \caption{\textit{Left panel:} Confidence areas of the halo mass $M_{\mathrm{h}}$ and concentration $c$ of central galaxies for the analysis of the EAGLE simulation using the one-dimensional method and the two-dimensional method, scaled with the input stellar-to-halo mass relation and the concentration--mass relation. \textit{Right panel:} Confidence areas of the halo mass $M_{\mathrm{h}}$ and concentration $c$ of satellite galaxies for the analysis of the EAGLE simulation using the one-dimensional method and the two-dimensional method, scaled with the input stellar-to-halo mass relation and the concentration--mass relation. The contours show the results of the maximum likelihood fit on  the central galaxies (red and purple) and satellite galaxies (orange and green). Crosses (in corresponding  colours) show the best-fitting values for each method and galaxy sample, and the red circles show the fiducial models. The contours are calculated from the contours obtained as a fit of the $f$ and $g$ parameters.}
        \label{fig:contours_eagle3}
\end{figure*}

We fit the two lens models\footnote{One for centrals and one for satellites, for a combined set of four parameters;  each model has a set of $(f, g)$ parameters.} to the mean tangential shear profiles per bin and  to the full ellipticity data using  Equations \ref{eq:tangential} and \ref{eq:max_like}. To account for the uncertainty in our ellipticity measurements we again assign the standard deviation of 0.3, scaled to the typical number density of GAMA and KiDS set-up due to the size of the pixel in our mass maps. This results in an uncertainty of $\sigma_{\epsilon} = 0.015$. The gain in precision is 3.9, which is the ratio of the area of the 68\% confidence level contours ($\,\mathrm{FoM}_{2\mathrm{D}} / \mathrm{FoM}_{1\mathrm{D}} = 3.9\,$). We show the separate credibility contours for the halo mass and concentration in Fig. \ref{fig:contours_eagle3} for the one-dimensional and two-dimensional methods, separated into contributions from central and satellite galaxies; both  values are scaled with the input stellar-to-halo mass relation and the concentration--mass relation, which then show the relative change of the halo masses and concentration from fiducial values obtained from the simulation.

\begin{table}
        \caption{Parameters of the stellar-to-halo mass relation.}
        \label{tab:smhm}
        \centering
        \begin{tabular}{lcccc} % four columns, alignment for each
                \toprule
                & $\alpha$ & $\beta$ & $\gamma$ & $c/c_{\mathrm{\,Duffy}}$\\ 
                \midrule
                Centrals & 11.81 & -0.68 & 8.69 & 0.6 \\ 
                Satellites & 11.72 & -0.79 & 9.44 & 0.25 \\
                \bottomrule
        \end{tabular}
        \tablefoot{Parameters of the stellar-to-halo mass relation  used in the analysis of the EAGLE simulation, used together with the functional form presented in Equation \ref{eq:smhm}, and the normalisations of the concentration--mass relation.}
\end{table}

\begin{table}
        \caption{Central values of the input and best-fit halo masses and concentrations.}
        \label{tab:cm}
        \centering
        \begin{tabular}{lcccc} % four columns, alignment for each
                \toprule
                & Input $M_{\text{h}}$ & Best-fit $M_{\text{h}}$ & Input $c$ & Best-fit $c$ \\ 
                \midrule
                Centrals & 7.17 & 6.62 & 5.62 & 6.08 \\ 
                Satellites & 1.33 & 1.30 & 2.72 & 2.96 \\ 
                \bottomrule
        \end{tabular}
        \tablefoot{Central values of the input and best-fit halo masses and concentrations for the $10^{10.8}$ to $10^{11.0} M_{\odot}$ stellar mass bin. The input values are a median of the halo masses and concentrations of haloes in that bin measured from the convergence fit to the EAGLE data and the output values are the predictions from the best-fit one-dimensional model, which can be seen in  Fig. \ref{fig:gamma_eagle}.}
\end{table}

At first sight it might seem that the results are somewhat biased with regard to the actual measured scaling relation of the EAGLE galaxies, but we do observe an almost equal effect on the $f$ and $g$ parameters for the two methods. This is not necessarily due to a bias in the analysis. After all, the intrinsic scatter in the concentration--mass relation and in the stellar mass-to-halo mass relation are not accounted for in the model and are the most likely cause of small shifts in the  methods presented. Both methods also give robust estimates for the properties of central and satellite galaxies and given the results, the two-dimensional method is much more precise in constraining the two fitted parameters than the one-dimensional method. Given the large uncertainty on the recovered parameters of the one-dimensional method for satellite galaxies, the one-dimensional method is unable to robustly capture the contribution from satellites in the KiDS+GAMA data, as was also demonstrated by \citet{Sifon2015}. Given the results of this exercise, we expect to capture the contribution from satellites when we apply the two-dimensional method to the data.

\section{Discussion and conclusions}
\label{sec:conclusions}

We have investigated the precision and bias of one and two-dimensional galaxy-galaxy lensing analyses of weak lensing data, using tangential averaged shear profiles and ellipticities, respectively, keeping in mind current and upcoming state-of-the-art large weak lensing galaxy surveys. The main difference between the two methods lies in the fact that the two-dimensional approach uses all the available information in an observed field. While the one-dimensional method uses only the ellipticities of source galaxies to infer the stacked tangential shear signal, the two-dimensional method uses actual relative positions of all the lens galaxies in a field and the ellipticities of all the sources in the field. Because the two-dimensional galaxy-galaxy lensing accounts for spatial configuration of the lens galaxies, the unique signatures in the shear field caused by overlapping regions of influence contain more information about the halo properties of the lenses we want to study and result in a significant improvement over the traditional one-dimensional stacking methods.

We  tested the method on mock observations generated in a semi-empirical way where we  assumed a model with the gravitational lenses represented by the NFW profiles with properties determined from observable quantities such as stellar mass, taking into account a typical configuration and properties of KiDS and GAMA surveys. We find that the two-dimensional method gives better constraints on those same parameters: the  FoM is  more than three times larger compared to the results from stacked tangential shear profiles. This suggests that there can be an equal amount of information hidden in the exact configuration of the lenses and their overlaps, which is lost when a one-dimensional method is used. The precision gain also depends on the lens density. In denser fields of gravitational lenses, the gain in precision from using the two-dimensional method is larger, as the signal becomes more heavily influenced by neighbouring gravitational lenses. We also studied the case where we removed a significant fraction of galaxies present in a mock field from our analysis, and while the two-dimensional method still gives us better constraints on the NFW parameters, the accuracy of these parameters starts to suffer because the modelling of the lenses does not account for the contributions of shears that are caused by the galaxies we left out of our analysis. While this indeed produces a noticeable bias, and thus needs to be corrected  to properly recover the true values of the parameters we study, the case where  such a large fraction of galaxies would be missed is rather severe. This effect of correlated structure---undetected galaxies that are clustered with the observed galaxies and the matter distribution on group scale---is in reality negligibly small \citep[as discussed in detail already by][]{Hudson1998}.

We  assumed a model where lenses are represented by the NFW profiles, up to constant pre-factors for the lensing signal amplitude and scale. We used the same lens model as well for the study on the EAGLE simulation \citep{Schaye2015, McAlpine2015}. As we  used the concentration--mass relation that closely describes the one measured in the EAGLE simulation and a stellar-to-halo mass relation of the EAGLE central and satellite galaxies in our lens model, we expected both methods to recover the input parameters values. We find that the two methods are able to almost perfectly recover these values,  and the small differences can be attributed to the non-ideal modelling of the galaxies in the EAGLE simulation. The two-dimensional method does indeed perform better.

Given  that the two-dimensional galaxy-galaxy lensing method requires knowledge of group (and/or cluster) membership, preferentially inferred from spectroscopic data, we identify two cases where using the two-dimensional method could be preferred over the one-dimensional method. The most obvious one is studying the group properties as a function of halo mass, where using the two-dimensional method can give better constraints on scaling relations of group halo mass with luminosity of central galaxies, their stellar mass, size, X-ray gas emission, and the concentration of such haloes. As we have precise  membership information of galaxies in clusters and because of the increased number density of these galaxies, the second case is to study the sub-halo mass function of galaxy clusters to a high precision. The two-dimensional galaxy-galaxy lensing, together with the group or cluster membership information that is available by using highly complete spectroscopic surveys is an obvious choice  for galaxy-galaxy studies on dense galaxy fields in general.

%%% ACKNOWLEDGEMENTS %%%
\begin{acknowledgements}
We thank the anonymous referee for the  very useful comments and suggestions. We thank Mike Hudson for the useful discussion on the topic. AD acknowledges support from grant number 614.001.541 and HH acknowledges support from Vici grant number 639.043.512, both financed by the Netherlands Organisation for Scientific Research (NWO). KK acknowledges support  from the Alexander von Humboldt Foundation. \\

This work has made use of Python (\url{http://www.python.org}), including the packages \texttt{numpy} (\url{http://www.numpy.org}) 
and \texttt{scipy} (\url{http://www.scipy.org}). The plots were produced with \texttt{matplotlib} \citep{Hunter2007}.
\end{acknowledgements}

\bibliographystyle{aa}
\bibliography{library}

\begin{thebibliography}{44}
\expandafter\ifx\csname natexlab\endcsname\relax\def\natexlab#1{#1}\fi

\bibitem[{Bartelmann \& Schneider(2001)}]{Bartelmann1999}
Bartelmann, M. \& Schneider, P. 2001, Phys. Rep., 340, 291

\bibitem[{Brouwer {et~al.}(2016)Brouwer, Cacciato, Dvornik, Eardley, Heymans,
  Hoekstra, Kuijken, McNaught-Roberts, Sif{\'{o}}n, Viola, Alpaslan, Bilicki,
  Bland-Hawthorn, Brough, Choi, Driver, Erben, Grado, Hildebrandt, Holwerda,
  Hopkins, de~Jong, Liske, McFarland, Nakajima, Napolitano, Norberg, Peacock,
  Radovich, Robotham, Schneider, Sikkema, van Uitert, {Verdoes Kleijn}, \&
  Valentijn}]{Brouwer2016}
Brouwer, M.~M., Cacciato, M., Dvornik, A., {et~al.} 2016, MNRAS, 462, 4451

\bibitem[{Brouwer {et~al.}(2017)Brouwer, Visser, Dvornik, Hoekstra, Kuijken,
  Valentijn, Bilicki, Blake, Brough, Buddelmeijer, Erben, Heymans, Hildebrandt,
  Holwerda, Hopkins, Klaes, Liske, Loveday, McFarland, Nakajima, Sif{\'{o}}n,
  \& Taylor}]{Brouwer2016a}
Brouwer, M.~M., Visser, M.~R., Dvornik, A., {et~al.} 2017, MNRAS, 466, 2547

\bibitem[{Cacciato {et~al.}(2014)Cacciato, van Uitert, \&
  Hoekstra}]{Cacciato2013}
Cacciato, M., van Uitert, E., \& Hoekstra, H. 2014, MNRAS, 437, 377

\bibitem[{Cooray \& Sheth(2002)}]{Cooray2002}
Cooray, A. \& Sheth, R. 2002, Phys. Rep., 372, 1

\bibitem[{Courteau {et~al.}(2014)Courteau, Cappellari, {De Jong}, Dutton,
  Emsellem, Hoekstra, Koopmans, Mamon, Maraston, Treu, \&
  Widrow}]{Courteau2014}
Courteau, S., Cappellari, M., {De Jong}, R.~S., {et~al.} 2014, Rev. Mod. Phys.,
  86, 47

\bibitem[{Crain {et~al.}(2015)Crain, Schaye, Bower, Furlong, Schaller, Theuns,
  {Dalla Vecchia}, Frenk, McCarthy, Helly, Jenkins, Rosas-Guevara, White, \&
  Trayford}]{Crain2015}
Crain, R.~A., Schaye, J., Bower, R.~G., {et~al.} 2015, MNRAS, 450, 1937

\bibitem[{de~Jong {et~al.}(2015)de~Jong, {Verdoes Kleijn}, Boxhoorn,
  Buddelmeijer, Capaccioli, Getman, Grado, Helmich, Huang, Irisarri, Kuijken,
  {La Barbera}, McFarland, Napolitano, Radovich, Sikkema, Valentijn, Begeman,
  Brescia, Cavuoti, Choi, Cordes, Covone, Dall'Ora, Hildebrandt, Longo,
  Nakajima, Paolillo, Puddu, Rifatto, Tortora, van Uitert, Buddendiek,
  Harnois-D{\'{e}}raps, Erben, Eriksen, Heymans, Hoekstra, Joachimi, Kitching,
  Klaes, Koopmans, K{\"{o}}hlinger, Roy, Sif{\'{o}}n, Schneider, Sutherland,
  Viola, \& Vriend}]{DeJong2015}
de~Jong, J. T.~A., {Verdoes Kleijn}, G.~A., Boxhoorn, D.~R., {et~al.} 2015,
  A{\&}A, 582, A62

\bibitem[{Driver {et~al.}(2011)Driver, Hill, Kelvin, Robotham, Liske, Norberg,
  Baldry, Bamford, Hopkins, Loveday, Peacock, Andrae, Bland-Hawthorn, Brough,
  Brown, Cameron, Ching, Colless, Conselice, Croom, Cross, {De Propris}, Dye,
  Drinkwater, Ellis, Graham, Grootes, Gunawardhana, Jones, van Kampen,
  Maraston, Nichol, Parkinson, Phillipps, Pimbblet, Popescu, Prescott,
  Roseboom, Sadler, Sansom, Sharp, Smith, Taylor, Thomas, Tuffs, Wijesinghe,
  Dunne, Frenk, Jarvis, Madore, Meyer, Seibert, Staveley-Smith, Sutherland, \&
  Warren}]{Driver2011}
Driver, S.~P., Hill, D.~T., Kelvin, L.~S., {et~al.} 2011, MNRAS, 413, 971

\bibitem[{Duffy {et~al.}(2008)Duffy, Schaye, Kay, \& {Dalla
  Vecchia}}]{Duffy2011}
Duffy, A.~R., Schaye, J., Kay, S.~T., \& {Dalla Vecchia}, C. 2008, MNRAS, 390,
  L64

\bibitem[{Dvornik {et~al.}(2018)Dvornik, Hoekstra, Kuijken, Schneider, Amon,
  Nakajima, Viola, Choi, Erben, Farrow, Heymans, Hildebrandt, Sif{\'{o}}n, \&
  Wang}]{Dvornik2018}
Dvornik, A., Hoekstra, H., Kuijken, K., {et~al.} 2018, MNRAS, 479, 1240

\bibitem[{Geiger \& Schneider(1999)}]{Geiger1999}
Geiger, B. \& Schneider, P. 1999, MNRAS, 302, 118

\bibitem[{Han {et~al.}(2015)Han, Eke, Frenk, Mandelbaum, Norberg, Schneider,
  Peacock, Jing, Baldry, Bland-Hawthorn, Brough, Brown, Liske, Loveday, \&
  Robotham}]{Han2014}
Han, J., Eke, V.~R., Frenk, C.~S., {et~al.} 2015, MNRAS, 446, 1356

\bibitem[{Hildebrandt {et~al.}(2017)Hildebrandt, Viola, Heymans, Joudaki,
  Kuijken, Blake, Erben, Joachimi, Klaes, Miller, Morrison, Nakajima, {Verdoes
  Kleijn}, Amon, Choi, Covone, de~Jong, Dvornik, {Fenech Conti}, Grado,
  Harnois-D{\'{e}}raps, Herbonnet, Hoekstra, K{\"{o}}hlinger, McFarland, Mead,
  Merten, Napolitano, Peacock, Radovich, Schneider, Simon, Valentijn, van~den
  Busch, van Uitert, \& {Van Waerbeke}}]{Hildebrandt2016}
Hildebrandt, H., Viola, M., Heymans, C., {et~al.} 2017, MNRAS, 465, 1454

\bibitem[{Hoekstra(2003)}]{Hoekstra2003a}
Hoekstra, H. 2003, MNRAS, 339, 1155

\bibitem[{Hoekstra(2014)}]{Hoekstra2013book}
Hoekstra, H. 2014, Proc. Int. Sch. Phys. Enrico Fermi, 186, 59

\bibitem[{Hoekstra {et~al.}(2003)Hoekstra, Franx, Kuijken, Carlberg, \&
  Yee}]{Hoekstra2003}
Hoekstra, H., Franx, M., Kuijken, K., Carlberg, R.~G., \& Yee, H. K.~C. 2003,
  MNRAS, 340, 609

\bibitem[{Hoekstra {et~al.}(2004)Hoekstra, Yee, \& Gladders}]{Hoekstra2004a}
Hoekstra, H., Yee, H. K.~C., \& Gladders, M.~D. 2004, ApJ, 606, 67

\bibitem[{Hudson {et~al.}(1998)Hudson, Gwyn, Dahle, \& Kaiser}]{Hudson1998}
Hudson, M.~J., Gwyn, S. D.~J., Dahle, H., \& Kaiser, N. 1998, ApJ, 503, 531

\bibitem[{Hunter(2007)}]{Hunter2007}
Hunter, J.~D. 2007, Comput. Sci. Eng., 9, 90

\bibitem[{Kaiser \& Squires(1993)}]{Kaiser1993}
Kaiser, N. \& Squires, G. 1993, ApJ, 404, 441

\bibitem[{Kuijken {et~al.}(2015)Kuijken, Heymans, Hildebrandt, Nakajima, Erben,
  de~Jong, Viola, Choi, Hoekstra, Miller, van Uitert, Amon, Blake, Brouwer,
  Buddendiek, Conti, Eriksen, Grado, Harnois-D{\'{e}}raps, Helmich, Herbonnet,
  Irisarri, Kitching, Klaes, {La Barbera}, Napolitano, Radovich, Schneider,
  Sif{\'{o}}n, Sikkema, Simon, Tudorica, Valentijn, {Verdoes Kleijn}, \& van
  Waerbeke}]{Kuijken2015}
Kuijken, K., Heymans, C., Hildebrandt, H., {et~al.} 2015, MNRAS, 454, 3500

\bibitem[{Leauthaud {et~al.}(2011)Leauthaud, Tinker, Behroozi, Busha, \&
  Wechsler}]{Leauthaud2011}
Leauthaud, A., Tinker, J., Behroozi, P.~S., Busha, M.~T., \& Wechsler, R.~H.
  2011, ApJ, 738, 45

\bibitem[{Liske {et~al.}(2015)Liske, Baldry, Driver, Tuffs, Alpaslan, Andrae,
  Brough, Cluver, Grootes, Gunawardhana, Kelvin, Loveday, Robotham, Taylor,
  Bamford, Bland-Hawthorn, Brown, Drinkwater, Hopkins, Meyer, Norberg, Peacock,
  Agius, Andrews, Bauer, Ching, Colless, Conselice, Croom, Davies, {De
  Propris}, Dunne, Eardley, Ellis, Foster, Frenk, H{\"{a}}u{\ss}ler, Holwerda,
  Howlett, Ibarra, Jarvis, Jones, Kafle, Lacey, Lange, Lara-L{\'{o}}pez,
  L{\'{o}}pez-S{\'{a}}nchez, Maddox, Madore, McNaught-Roberts, Moffett, Nichol,
  Owers, Palamara, Penny, Phillipps, Pimbblet, Popescu, Prescott, Proctor,
  Sadler, Sansom, Seibert, Sharp, Sutherland, V{\'{a}}zquez-Mata, van Kampen,
  Wilkins, Williams, \& Wright}]{Liske2015}
Liske, J., Baldry, I.~K., Driver, S.~P., {et~al.} 2015, MNRAS, 452, 2087

\bibitem[{Matthee {et~al.}(2017)Matthee, Schaye, Crain, Schaller, Bower, \&
  Theuns}]{Matthee2016}
Matthee, J., Schaye, J., Crain, R.~A., {et~al.} 2017, MNRAS, 465, 2381

\bibitem[{McAlpine {et~al.}(2016)McAlpine, Helly, Schaller, Trayford, Qu,
  Furlong, Bower, Crain, Schaye, Theuns, {Dalla Vecchia}, Frenk, McCarthy,
  Jenkins, Rosas-Guevara, White, Baes, Camps, \& Lemson}]{McAlpine2015}
McAlpine, S., Helly, J., Schaller, M., {et~al.} 2016, A{\&}C, 15, 72

\bibitem[{McKay {et~al.}(1979)McKay, Beckman, \& Canover}]{McKay1979}
McKay, M., Beckman, R., \& Canover, W. 1979, Technometrics, 21, 239

\bibitem[{Navarro {et~al.}(1996)Navarro, Frenk, \& White}]{Navarro1995}
Navarro, J.~F., Frenk, C.~S., \& White, S. D.~M. 1996, ApJ, 462, 563

\bibitem[{Peacock \& Smith(2000)}]{Peacock2000}
Peacock, J.~A. \& Smith, R.~E. 2000, MNRAS, 318, 1144

\bibitem[{{Planck Collaboration} {et~al.}(2013){Planck Collaboration}, Ade,
  Aghanim, Armitage-Caplan, Arnaud, Ashdown, Atrio-Barandela, Aumont,
  Baccigalupi, Banday, Barreiro, Bartlett, Battaner, Benabed, Beno{\^{i}}t,
  Benoit-L{\'{e}}vy, Bernard, Bersanelli, Bielewicz, Bobin, Bock, Bonaldi,
  Bond, Borrill, Bouchet, Bridges, Bucher, Burigana, Butler, Calabrese,
  Cappellini, Cardoso, Catalano, Challinor, Chamballu, Chary, Chen, Chiang,
  Chiang, Christensen, Church, Clements, Colombi, Colombo, Couchot, Coulais,
  Crill, Curto, Cuttaia, Danese, Davies, Davis, de~Bernardis, de~Rosa,
  de~Zotti, Delabrouille, Delouis, D{\'{e}}sert, Dickinson, Diego, Dolag, Dole,
  Donzelli, Dor{\'{e}}, Douspis, Dunkley, Dupac, Efstathiou, Elsner,
  En{\ss}lin, Eriksen, Finelli, Forni, Frailis, Fraisse, Franceschi, Gaier,
  Galeotta, Galli, Ganga, Giard, Giardino, Giraud-H{\'{e}}raud, Gjerl{\o}w,
  Gonz{\'{a}}lez-Nuevo, G{\'{o}}rski, Gratton, Gregorio, Gruppuso, Gudmundsson,
  Haissinski, Hamann, Hansen, Hanson, Harrison, Henrot-Versill{\'{e}},
  Hern{\'{a}}ndez-Monteagudo, Herranz, Hildebrandt, Hivon, Hobson, Holmes,
  Hornstrup, Hou, Hovest, Huffenberger, Jaffe, Jaffe, Jewell, Jones, Juvela,
  Keih{\"{a}}nen, Keskitalo, Kisner, Kneissl, Knoche, Knox, Kunz, Kurki-Suonio,
  Lagache, L{\"{a}}hteenm{\"{a}}ki, Lamarre, Lasenby, Lattanzi, Laureijs,
  Lawrence, Leach, Leahy, Leonardi, Le{\'{o}}n-Tavares, Lesgourgues, Lewis,
  Liguori, Lilje, Linden-V{\o}rnle, L{\'{o}}pez-Caniego, Lubin,
  Mac{\'{i}}as-P{\'{e}}rez, Maffei, Maino, Mandolesi, Maris, Marshall, Martin,
  Mart{\'{i}}nez-Gonz{\'{a}}lez, Masi, Massardi, Matarrese, Matthai, Mazzotta,
  Meinhold, Melchiorri, Melin, Mendes, Menegoni, Mennella, Migliaccio, Millea,
  Mitra, Miville-Desch{\^{e}}nes, Moneti, Montier, Morgante, Mortlock, Moss,
  Munshi, Murphy, Naselsky, Nati, Natoli, Netterfield, N{\o}rgaard-Nielsen,
  Noviello, Novikov, Novikov, O'Dwyer, Osborne, Oxborrow, Paci, Pagano, Pajot,
  Paoletti, Partridge, Pasian, Patanchon, Pearson, Pearson, Peiris, Perdereau,
  Perotto, Perrotta, Pettorino, Piacentini, Piat, Pierpaoli, Pietrobon,
  Plaszczynski, Platania, Pointecouteau, Polenta, Ponthieu, Popa, Poutanen,
  Pratt, Pr{\'{e}}zeau, Prunet, Puget, Rachen, Reach, Rebolo, Reinecke,
  Remazeilles, Renault, Ricciardi, Riller, Ristorcelli, Rocha, Rosset, Roudier,
  Rowan-Robinson, Rubi{\~{n}}o-Mart{\'{i}}n, Rusholme, Sandri, Santos,
  Savelainen, Savini, Scott, Seiffert, Shellard, Spencer, Starck, Stolyarov,
  Stompor, Sudiwala, Sunyaev, Sureau, Sutton, Suur-Uski, Sygnet, Tauber,
  Tavagnacco, Terenzi, Toffolatti, Tomasi, Tristram, Tucci, Tuovinen,
  T{\"{u}}rler, Umana, Valenziano, Valiviita, {Van Tent}, Vielva, Villa,
  Vittorio, Wade, Wandelt, Wehus, White, White, Wilkinson, Yvon, Zacchei, \&
  Zonca}]{PlanckCollaboration2014}
{Planck Collaboration}, Ade, P. a.~R., Aghanim, N., {et~al.} 2013, A{\&}A, 571,
  A16

\bibitem[{Robotham {et~al.}(2011)Robotham, Norberg, Driver, Baldry, Bamford,
  Hopkins, Liske, Loveday, Merson, Peacock, Brough, Cameron, Conselice, Croom,
  Frenk, Gunawardhana, Hill, Jones, Kelvin, Kuijken, Nichol, Parkinson,
  Pimbblet, Phillipps, Popescu, Prescott, Sharp, Sutherland, Taylor, Thomas,
  Tuffs, van Kampen, \& Wijesinghe}]{Robotham2011}
Robotham, A. S.~G., Norberg, P., Driver, S.~P., {et~al.} 2011, MNRAS, 416, 2640

\bibitem[{Schaller {et~al.}(2015)Schaller, Frenk, Bower, Theuns, Jenkins,
  Schaye, Crain, Furlong, {Dalla Vecchia}, \& McCarthy}]{Schaller2015}
Schaller, M., Frenk, C.~S., Bower, R.~G., {et~al.} 2015, MNRAS, 451, 1247

\bibitem[{Schaye {et~al.}(2015)Schaye, Crain, Bower, Furlong, Schaller, Theuns,
  {Dalla Vecchia}, Frenk, McCarthy, Helly, Jenkins, Rosas-Guevara, White, Baes,
  Booth, Camps, Navarro, Qu, Rahmati, Sawala, Thomas, \& Trayford}]{Schaye2015}
Schaye, J., Crain, R.~A., Bower, R.~G., {et~al.} 2015, MNRAS, 446, 521

\bibitem[{Schneider(2003)}]{Schneider2003}
Schneider, P. 2003, ArXiv e-prints [\eprint[arXiv]{0306465}]

\bibitem[{Schneider \& Rix(1997)}]{Schneider1997}
Schneider, P. \& Rix, H.-W. 1997, ApJ, 474, 25

\bibitem[{Seljak(2000)}]{Seljak2000}
Seljak, U. 2000, MNRAS, 318, 203

\bibitem[{Sif{\'{o}}n {et~al.}(2015)Sif{\'{o}}n, Cacciato, Hoekstra, Brouwer,
  van Uitert, Viola, Baldry, Brough, Brown, Choi, Driver, Erben, Grado,
  Heymans, Hildebrandt, Joachimi, de~Jong, Kuijken, McFarland, Miller,
  Nakajima, Napolitano, Norberg, Robotham, Schneider, \& Kleijn}]{Sifon2015}
Sif{\'{o}}n, C., Cacciato, M., Hoekstra, H., {et~al.} 2015, MNRAS, 454, 3938

\bibitem[{Sonnenfeld \& Leauthaud(2018)}]{Sonnenfeld2017}
Sonnenfeld, A. \& Leauthaud, A. 2018, MNRAS, 477, 5460

\bibitem[{van Uitert {et~al.}(2016)van Uitert, Cacciato, Hoekstra, Brouwer,
  Sif{\'{o}}n, Viola, Baldry, Bland-Hawthorn, Brough, Brown, Choi, Driver,
  Erben, Heymans, Hildebrandt, Joachimi, Kuijken, Liske, Loveday, McFarland,
  Miller, Nakajima, Peacock, Radovich, Robotham, Schneider, Sikkema, Taylor, \&
  {Verdoes Kleijn}}]{vanUitert2016}
van Uitert, E., Cacciato, M., Hoekstra, H., {et~al.} 2016, MNRAS, 459, 3251

\bibitem[{van Uitert {et~al.}(2011)van Uitert, Hoekstra, Velander, Gilbank,
  Gladders, \& Yee}]{VanUitert2011}
van Uitert, E., Hoekstra, H., Velander, M., {et~al.} 2011, A{\&}A, 534, A14

\bibitem[{Velander {et~al.}(2014)Velander, van Uitert, Hoekstra, Coupon, Erben,
  Heymans, Hildebrandt, Kitching, Mellier, Miller, {Van Waerbeke}, Bonnett, Fu,
  Giodini, Hudson, Kuijken, Rowe, Schrabback, \& Semboloni}]{Velander2014}
Velander, M., van Uitert, E., Hoekstra, H., {et~al.} 2014, MNRAS, 437, 2111

\bibitem[{Velliscig {et~al.}(2017)Velliscig, Cacciato, Hoekstra, Schaye,
  Heymans, Hildebrandt, Loveday, Norberg, Sif{\'{o}}n, Schneider, van Uitert,
  Viola, Brough, Erben, Holwerda, Hopkins, \& Kuijken}]{Velliscig2016}
Velliscig, M., Cacciato, M., Hoekstra, H., {et~al.} 2017, MNRAS, 471, 2856

\bibitem[{Viola {et~al.}(2015)Viola, Cacciato, Brouwer, Kuijken, Hoekstra,
  Norberg, Robotham, van Uitert, Alpaslan, Baldry, Choi, de~Jong, Driver,
  Erben, Grado, Graham, Heymans, Hildebrandt, Hopkins, Irisarri, Joachimi,
  Loveday, Miller, Nakajima, Schneider, Sif{\'{o}}n, \& {Verdoes
  Kleijn}}]{Viola2015}
Viola, M., Cacciato, M., Brouwer, M., {et~al.} 2015, MNRAS, 452, 3529

\bibitem[{Wright \& Brainerd(2000)}]{Wright1999}
Wright, C.~O. \& Brainerd, T.~G. 2000, ApJ, 534, 34

\end{thebibliography}

\end{document}